\documentclass[journal]{IEEEtran}
\usepackage{url,amsmath,amssymb,subfigure,epsfig,color,array,cases,paralist,graphicx,times,cite,bm,amsthm,multirow,stfloats}

\IEEEoverridecommandlockouts
\newcommand{\pr}{\text{P}}
\newcommand{\E}{\text{E}}
\newcommand{\ed}{\stackrel{\text{d}}{=}}
\newcommand{\deq}{\triangleq}

\newtheorem{definition}{Definition}
\newtheorem{theorem}{Theorem}
\newtheorem{lemma}{Lemma}
\newtheorem{remark}{Remark}

\makeatletter
\def\ps@headings{%
\def\@oddhead{\mbox{}\scriptsize\rightmark \hfil \thepage}%
\def\@evenhead{\scriptsize\thepage \hfil \leftmark\mbox{}}%
\def\@oddfoot{}%
\def\@evenfoot{}}
\makeatother
\begin{document}
\title{Providing Probabilistic Guarantees on the Time of Information Spread in Opportunistic Networks}
\author{Yoora Kim$^\dag$, Kyunghan Lee$^\ddag$, Ness B. Shroff$^\dag$, and Injong Rhee$^\sharp$ \\
$^\dag$\{kimy, shroff\}@ece.osu.edu, $^\ddag$khlee@unist.ac.kr, $^\sharp$rhee@ncsu.edu

\thanks{This work has been supported in part by the National Science Foundation CNS-1065136, CNS-0910868, and CNS-1016216; the Army Research Office MURI grants W911NF-08-1-0238 and W911NF-12-1-0385; and the year of 2012 Research Fund of the UNIST (Ulsan National Institute of Science and Technology).}
\thanks{Y. Kim is with the Department of Electrical and Computer Engineering, The Ohio State University, USA. K. Lee is with the School of Electrical and Computer Engineering, UNIST, Korea. N. B. Shroff is with the Department of Electrical and Computer Engineering and the Department of Computer Science and Engineering, The Ohio State University, USA. I. Rhee is with the Department of Computer Science, North Carolina State University, USA.}}
\maketitle

\begin{abstract}
A variety of mathematical tools have been developed for predicting the spreading patterns in a number of varied environments including infectious diseases, computer viruses, and urgent messages broadcast to mobile agents (e.g., humans, vehicles, and mobile devices). These tools have mainly focused on estimating the average time for the spread to reach a fraction (e.g., $\alpha$) of the agents, i.e., the so-called average completion time $E(T_{\alpha})$. We claim that providing probabilistic guarantee on the time for the spread $T_{\alpha}$ rather than only its average gives a much better understanding of the spread, and hence could be used to design improved methods to prevent epidemics or devise accelerated methods for distributing data. To demonstrate the benefits, we introduce a new metric $G_{\alpha, \beta}$ that denotes the time required to guarantee $\alpha$ completion with probability $\beta$, and develop a new framework to characterize the distribution of $T_\alpha$ for various spread parameters such as number of seeds, level of contact rates, and heterogeneity in contact rates. We apply our technique to an experimental mobility trace of taxies in Shanghai and show that our framework enables us to allocate resources (i.e., to control spread parameters) for acceleration of spread in a far more efficient way than the state-of-the-art.
\end{abstract}

\section{Introduction}\label{sec:intro}
Spreading patterns of pandemics~\cite{keeling01}, computer viruses~\cite{Sellke08}, and information~\cite{zhang07, jacquet10} have been widely studied in various research disciplines including epidemics, biology, physics, sociology, and computer networks. In these disciplines, most studies have been devoted to characterizing spread behaviors toward a network of mobile agents including humans, vehicles, and mobile devices\footnote{We will interchangeably use agents and nodes unless confusion arises.} over time. These studies can be classified into two groups based on their objectives. Interestingly, both these objects lie in opposite directions: slowing down or acceleration of spread.
For the research that deals with biological and electronic viruses, how to slow down the spread has been the most important question to be answered. On the other hand, another set of research work for computer data and information distribution has pursued designing engineering methods to accelerate the spread.

Whatever the goals are, existing studies have relied on common mathematical frameworks such as the branching process, mean-field approximation, and stochastic differential equations~\cite{andersson00}. Due to the characteristics of these frameworks, the spread of virus or information has generally been analyzed in terms of its average behavior under various epidemic models summarized in \cite{keeling05}, where epidemic models define whether agents are recoverable\footnote{A virus that cannot be recovered can be considered to be identical to undeletable or unforgettable information.} or not and whether they become immune after recovery or are still susceptible to infection. Here, average behavior typically indicates $\E[N_t]$ where $N_t$ denotes the number of infected nodes in the network at time~$t$.

Average analysis gives an answer to a question on how many nodes are infected (or informed) \textit{on average} under a specific epidemic model after a time duration $t$ from the emergence of a virus (or generation of information). There have been many extensions to this analysis through aforementioned frameworks. The authors of \cite{karsai11} identified how much a network topology affects the speed of virus spreading and the authors of \cite{piet09} derived a closed form equation of the critical level of virus infection rate allowing a virus to persist in a network when the virus is recoverable with a certain rate. More realistic average spread behaviors of a virus with the heterogeneity inherent in human mobility patterns have been studied through simulations in~\cite{yang11}. In computer networks, \cite{zou02} analyzed the average propagation behavior of code red worm in the Internet using measurement data from ISP and an epidemic model. \cite{zhang07} applied understanding on the average behavior of virus spread to information propagation in delay tolerant  networks. Similarly, \cite{Sellke08} analyzed the average spread behaviors of self-propagating worms on the Internet using branching process.

While there has been a plethora of work on average analysis, the problem of allocating optimal amounts of resource to a network of a set of nodes for slowing down or accelerating spread has been under-explored.\footnote{\cite{optimalmsn09} studied the optimal allocation of wireless channels of a carrier to mobile nodes in a content delivery network, which maximizes the sum utility defined with the content delivery time to the nodes. In the work, a bound on the content delivery time was studied, but its exact distribution was left unsolved.} Specifically, higher order spread behaviors over time rather than average behaviors to design optimal resource allocation have not been well understood. The right question should be what will be the distribution of the number of infected nodes at time $t$, which is equivalent to what will be the temporal distribution of the event that $n$ nodes are infected. Characterizing the temporal distribution of spread allows one to guarantee the time for spread with high probability and it leads to control knobs for allocating resources to a network with its own purpose of spread. However, understanding the temporal distribution involves non-trivial challenges since there is a huge dimension of diversity in contact events among nodes in a network. 

In order to address the challenges involved in obtaining deeper understanding of resource allocation, in this paper, we propose a new analytical framework based on CTMC (continuous time Markov chain), which allows us to fully characterize the temporal aspect of spread behaviors. For simplicity, we put our emphasis on information distribution among intermittently meeting mobile nodes forming an opportunistic network, i.e., a mobile social network, but our results are easily applicable to general spread of epidemics. Our framework is capable of answering many intriguing engineering questions such as ``what is the distribution of time for a network to have 75\% penetration rate?" and ``If 75\% penetration is aimed, when is the time to guarantee that level of penetration with 99\% of confidence?". It can also answer a more fundamental question involving heterogeneity of nodes in a network, ``Does heterogeneity help or hurt spreading?" We show the efficacy of our solution in answering these questions with the use of one of the largest experimental GPS (global positioning system) trace of taxies in Shanghai. Our simulation studies on the trace provide added verification that our framework is robust and enables us to engineer the network in a far more efficient way than existing understandings of spread.


The rest of the paper is organized as follows. In Section~\ref{sec:model}, we provide our system model along with definitions of relevant metrics. In Section~\ref{sec:framework}, we develop our analytical framework and present major analytical results. Based on our framework, we characterize the temporal distribution of spread behavior and provide their applications in Section~\ref{sec:characterization}. We present simulation studies using Shanghai taxi trace and conclude our paper in Sections~\ref{sec:numerical} and~\ref{sec:conclusion}, respectively. 
\section{Model Description}\label{sec:model}
\subsection{Overview of Epidemic Models}
In classic epidemiology, an individual (i.e., node) is classified into either susceptible, infected, or removed (or immune) according to its status for a disease~\cite{andersson00}. A susceptible individual refers to the one who is not infected yet, but is prone to be infected. An infected individual refers to the one who already got the disease and is capable of spreading it to susceptible individuals. A removed individual indicates the one who was previously infected but became immune to the disease. These three classifications are conventionally denoted by S, I, and R, respectively, and induce SIS, SIR, and SI epidemic models and their variants. In this paper, we focus on the SI model in which once a susceptible individual is infected, it stays infected for the remainder of the epidemic process. The SI model fits particularly well with information spread in opportunistic networks since once a data is delivered to an individual, it is considered that the data is delivered to its upper layer and it is no longer required (i.e., permanently infected).

\subsection{Our System Model}
We consider a network (or a population) consisting of~$N$ mobile nodes. 
We assume that all nodes in the network can be classified into~$K$ different types according to their mobility patterns and epidemic attributes. We denote the collection of the $k$th type of nodes as group~$k\,(k=1,\ldots,K)$. Let $N_k$ be the number of nodes in group~$k$ and denote $\bm{N}\deq (N_k)_{1\le k\le K}$. Then, we have $|\bm{N}| \deq \sum_{k}N_k = N$ (throughout this paper, we use a bold font symbol for an arbitrary vector or a matrix notation. In addition, for a vector $\bm{V}=(V_k)$, we define the operation $|\bm{V}|$ as $|\bm{V}| \deq \sum_{k}V_k$).

In our model, the mechanism of information (or a packet or a virus) spread is as follows: initially, the information is delivered to some selected nodes, which we call \emph{seeds}.\footnote{Note that being selected as seeds can be of willing or unwilling. For instance, a seed of a virus gets the virus unwillingly.} Whenever a seed, say node~$a$, meets a susceptible node not having the information yet, it spreads the information to the susceptible node with probability $\varphi_a\in(0,1]$. Then, the susceptible node, say node~$b$, receives the information successfully with probability $\psi_b\in(0,1]$ and becomes infected (or informed). Once the susceptible node becomes infected, it stays infected for the remainder of the spreading process, and is involved in disseminating the information in a similar manner as the seed. The spreading process ends when all nodes in the network obtain the information. In our spreading model, the probabilities $\varphi_a$ and $\psi_b$ can be interpreted as the infectivity and the susceptibility of nodes~$a$ and $b$, respectively. For instance, in the case of rumor propagation, $\varphi_a$ quantifies the tendency of person~$a$ to gossip, while $\psi_b$ quantifies the receptive nature of a listener~$b$ to the rumor. For the case of packet dissemination in an opportunistic network, $\varphi_a$ represents the probability that node $a$ schedules to transmit a packet, and $\psi_b$ represents the probability of successful packet reception at node~$b$, which depends on, e.g., the contact period, number of contending nodes, and channel conditions.

The stochastic characteristic of a pairwise meeting process is a critical factor that determines the temporal behavior of the spreading process. In the literature, it has been recently shown that the time duration between two consecutive contacts of a pair of nodes, called \emph{pairwise inter-contact time}, can be modeled by an exponential random variable, e.g.,~\cite{conan07, gao09, kim12tr}. In~\cite{conan07}, exponential inter-contact patterns are validated experimentally using three different mobility data sets. When nodes follow L\'{e}vy flight mobility, which is known to closely mimic human mobility patterns~\cite{lee:slaw09}, the authors in~\cite{kim12tr} mathematically proved that the inter-contact time distribution is bounded by an exponential distribution. Thus, in this paper we assume that the pairwise inter-contact time between nodes~$a$ and~$b$, denoted by $M_{a,b}$, follows an exponential distribution with rate $\lambda_{a,b}\,(>0)$, i.e.,
\begin{align}\label{eqn:exp meeting time}
\pr\!\left\{M_{a,b} > t \right\} = \exp\!\left(-\lambda_{a,b} t\right), \quad t \ge 0.
\end{align}
Suppose that node~$a$ is infected and node~$b$ is susceptible. From the meeting process between them, we can obtain the infection time $M^{\text{eff}}_{a,b}$ by taking the infectivity~$\varphi_a$ and the susceptibility~$\psi_b$ into account. From~(\ref{eqn:exp meeting time}), we have:
\begin{align}\label{eqn:exp meeting time:eff}
\begin{split}
\pr\!\left\{M^{\text{eff}}_{a,b} > t \right\} &= \exp\!\left(-\lambda_{a,b}\varphi_{a} \psi_{b} t\right), \quad t \ge 0.
\end{split}
\end{align}
That is, the infection rate $\lambda_{a,b}^\text{eff}$ between an infected node~$a$ and a susceptible node~$b$ becomes $\lambda_{a,b}^\text{eff}= \lambda_{a,b}\varphi_{a} \psi_{b}$. The detailed proof of~(\ref{eqn:exp meeting time:eff}) is given in~Appendix~\ref{sec:appendix:proof of eff meeting time}. Since all nodes in the same group are stochastically identical in terms of mobility pattern and epidemic attribute, the rate $\lambda_{a,b}^{\text{eff}}$ should be determined by the group indices of nodes~$a$ and~$b$. Thus, we can rewrite the infection rate as $\lambda_{a,b}^\text{eff} = \lambda^\star_{\mathcal{F}(a),\mathcal{F}(b)}$, where the subscripts $\mathcal{F}(a)$ and $\mathcal{F}(b)$ denote the group indices of node $a$ and $b$, respectively. For later use, we define a rate matrix $\bm{\Lambda}$ as
\begin{align*}
\bm{\Lambda}\deq \big(\lambda^\star_{k_1,k_2}\big)_{1\le k_1,k_2 \le K}.
\end{align*}

Our spreading model is general in that it covers a variety of scenarios from homogeneous to totally heterogeneous cases. For instance, when $K=1$ our model reduces to the homogeneous case where all nodes in the network are identical with the same infection rate $\lambda^\star_{\mathcal{F}(a),\mathcal{F}(b)} = \lambda_{1,1}^\star\,(\deq \lambda^\star)$ for any $a,b$. On the other hand, when $K=N$ our model induces a totally heterogeneous case where each individual node uniquely forms a group. When $K = 2,\ldots,N-1$, our model is able to capture heterogeneity arising from multiple communities. In addition to heterogeneity, our model is capable of characterizing the impact of various spread parameters such as the level of contact rates and population size by varying the values of the rate matrix~$\bm{\Lambda}$ and group sizes $\bm{N}$.

\subsection{Performance Metrics}
In this subsection, we describe our performance metrics in detail. Let $S_k(t)$ be the number of susceptible nodes in group~$k$ at time~$t\,(\ge 0)$. Let $I_k(t)$ be the number of infected nodes in group $k$ at time~$t$. Then, we have $S_k(t) + I_k(t) = N_k$ for all $k$ and $t$. The key performance metric of our interest is the $\alpha$-completion time as defined below: 

\begin{definition}[$\alpha$-completion time]\label{def:alpha completion time} For $\alpha\in (0,1]$, the time required for infecting $\alpha$~fraction of the total population, denoted by $T_\alpha$, is given by:
\begin{align}\label{eqn:def:alpha completion time}
T_\alpha \deq \inf \bigg\{t: \sum_{k=1}^{K}I_k(t) \ge \alpha N\bigg\}.
\end{align}
We call $T_\alpha$ the \emph{$\alpha$-completion time} throughout this paper.
\end{definition}

$T_{\alpha}$ has a strong connection with existing studies that have characterized the average number of infected nodes at time $t$ (i.e., $\sum_k \E[I_k (t)]$) using various mathematical tools, because $\E[T_\alpha]$ is a dual of $\sum_k \E[I_k (t)]$. However, to better understand the spread behavior and to better design spread prevention or acceleration methods, characterization of the distribution of $T_\alpha$ beyond simply the mean is needed. To this end, we introduce a new metric, called $(\alpha,\beta)$-guaranteed time, as defined next:

\begin{definition}[$(\alpha,\beta)$-guaranteed time]\label{def:guaranteed time} For $\alpha\in (0,1]$ and $\beta \in(0,1)$, the minimum time required to guarantee spread to $\alpha$ fraction of the total population with probability~$\beta$, denoted by $G_{\alpha,\beta}$, is defined by:
\begin{align}\label{eqn:def:guaranteed time}
G_{\alpha,\beta} \deq \inf \big\{t: \pr\{T_\alpha > t\} \le 1-\beta\big\}.
\end{align}
We call $G_{\alpha,\beta}$ the \emph{$(\alpha,\beta)$-guaranteed time} throughout this paper.
\end{definition}

Note that the probability $1-\beta$ in~(\ref{eqn:def:guaranteed time}) can be interpreted as an outage probability that the actual spread time~$T_\alpha$ exceeds the guaranteed time~$G_{\alpha,\beta}$. In this sense,~$G_{\alpha,\beta}$ can be used to predict the range of spread time and the confidence of the prediction: the higher we set the value of~$\beta$, the greater the confidence in the prediction. Thus, $G_{\alpha, \beta}$ facilitates avoiding underestimating the required resources for spreading information to a network. The ratio $R_{\alpha, \beta}$ defined below describes just how much $\E[T_\alpha]$ underestimates the spread time compared to the guaranteed time.

\begin{definition}[$(\alpha,\beta)$-average to guaranteed time ratio]\label{def:avg to guaranteed completion ratio} For $\alpha\in (0,1]$ and $\beta \in (0,1)$, the ratio $R_{\alpha,\beta}$ is defined by
\begin{align}\label{eqn:def:avg to guaranteed completion ratio}
R_{\alpha,\beta} \deq \frac{G_{\alpha,\beta}}{\E[T_\alpha]}.
\end{align}
We call $R_{\alpha,\beta}$ the \emph{$(\alpha,\beta)$-average to guaranteed time ratio} throughout this paper.
\end{definition}

Finally, we define the set of seeds in each group. Let $s_k \deq I_k(0)$ be the number of seeds in group~$k$. If $\sum_{k}s_k \ge \alpha N$, then we have a trivial result of $T_\alpha = 0$, $G_{\alpha,\beta} = 0$, and $R_{\alpha,\beta} = 1$ for any $\beta\in(0,1)$. Therefore, in the rest of the paper, we only consider the regime of $\sum_{k}s_k < \alpha N$.

\section{Basic Temporal Analysis Framework}\label{sec:framework}
In this section, we develop an analytical framework for deriving the performance metrics defined in~(\ref{eqn:def:alpha completion time}),~(\ref{eqn:def:guaranteed time}), and~(\ref{eqn:def:avg to guaranteed completion ratio}). We use the following three steps in our analysis: first, we identify the temporal behavior of the total number of infected nodes $\{\sum_{k}I_k(t);t\ge 0\}$ (See Lemma~\ref{lemma:CTMC model}). Using the result in Lemma~\ref{lemma:CTMC model}, we are able to obtain the distribution of the $\alpha$-completion time~$T_\alpha$ (See Lemma~\ref{lemma:distribution of completion time}). Finally, we give formulas for our performance metrics (See Lemma~\ref{lemma:formula for metrics}).

\smallskip
\noindent\emph{Step 1:} According to Definition~\ref{def:alpha completion time}, we need the temporal distribution of the total number of infected nodes $\sum_{k}I_k(t)$. Directly solving it appears to be intractable (as illustrated in the following Example~II for the case $K=2$). However, we prove that the \emph{joint temporal distribution} of $I_k(t)$ can be derived from the theory of multi-dimensional CTMC, which could be also used to identify the distribution of $\sum_{k}I_k(t)$. The result is summarized in Lemma~\ref{lemma:CTMC model} and its derivation is explained through the following Examples~I and~II.

\begin{lemma}[CTMC model]\label{lemma:CTMC model} For $K=1,2,\ldots$, let
\begin{align*}
\bm{I}(t) \deq \big(I_1(t),I_2(t),\ldots,I_K(t)\big).
\end{align*}
Then, the process $\{\bm{I}(t);t\ge 0\}$ is a $K$-dimensional CTMC. Further, it has the following properties:
\begin{enumerate}[(P1)]
\item The state space is given by~$\mathcal{E}\deq \prod_{k=1}^{K}\{0,\ldots,N_k\}\setminus\bm{0}$ and is decomposed into transient state space~$\mathcal{E}^\star$ and absorbing state space~$\mathcal{E}^o$ as:
    \begin{align*}
    \mathcal{E}^\star &\deq \{\bm{e}\in\mathcal{E}:|\bm{e}| <|\bm{N}|\}, \\
    \mathcal{E}^o &\deq \{\bm{N} = (N_1,\ldots,N_K)\}.
    \end{align*}
    Without loss of generality, we assume that the states in $\mathcal{E}=\{\bm{e}_1, \bm{e}_2, \ldots\}$ are arranged as $|\bm{e}_1| \le |\bm{e}_2|\le \cdots$.
\item By the property~(P1), the infinitesimal generator $\bm{Q}$ of the Markov chain is of the following form:
    \begin{align*}
    \bm{Q} = \left[\begin{matrix}
    \bm{F} & \bm{F}^o \\
    \bm{0} & 0
    \end{matrix}\right],
    \end{align*}
    where $\bm{F}=(F_{i,j})$ is a matrix representing transition rate from $\mathcal{E}^\star$ to $\mathcal{E}^\star$, and $\bm{F}^o$ is a column vector representing transition rate from $\mathcal{E}^\star$ to $\mathcal{E}^o$. Due to its importance, we call the matrix~$\bm{F}$ the \emph{fundamental matrix}.
\item Assume $\pr\{\bm{I}(0)\in\mathcal{E}^\star\}=1$. For a given time $t>0$, let $\bm{\pi}(t) \deq (\pr\{\bm{I}(t)=\bm{e}\})_{\bm{e}\in\mathcal{E}^\star}$ be the distribution of $\bm{I}(t)$ on $\mathcal{E}^\star$. Then, it is determined from its initial distribution $\bm{\pi}(0)$ and the fundamental matrix~$\bm{F}$ as~\cite{bremaud08}:
    \begin{align*}
    \bm{\pi}(t) = \bm{\pi}(0) \exp(\bm{F}t).
    \end{align*}
    The distribution of $\bm{I}(t)$ on $\mathcal{E}^o$ is then obtained by $\pr\{\bm{I}(t) = \bm{N}\} = 1 - |\bm{\pi}(t)|$.
\item Let the $i$th and the $j$th states in $\mathcal{E}$ be denoted by $\bm{e}_i = (i_k)_{1\le k\le K}$ and $\bm{e}_j = (j_k)_{1\le k\le K}$, respectively. Then, $\bm{Q} = (Q_{i,j})$ is obtained as:
    \begin{align*}
    Q_{i,j} &=
    \begin{cases}\sum_{l}I_{(\bm{e}_i, \bm{e}_j):l}(N_l-i_l)\sum_{k}i_k \lambda^\star_{k,l} & \text{if } i \neq j, \\
    -\sum_{l\neq i} Q_{i,l} & \text{if } i=j,
    \end{cases}
    \end{align*}
    where $I_{(\bm{e}_i, \bm{e}_j):l} \in\{0,1\}$ and takes 1 if and only if $j_l = i_l+1$ and $j_k = i_k$ for all $k\neq l$. Then, we can obtain $\bm{F}$ by restricting $\bm{Q}$ to the space $\mathcal{E}^\star$ as $\bm{F} = \bm{Q}|_{\mathcal{E}^\star \times \mathcal{E}^\star}$, i.e., $\bm{F} = (Q_{i,j})$ for all $i, j$ such that $\bm{e}_i, \bm{e}_j \in\mathcal{E}^\star$.
\end{enumerate}
\end{lemma}

\noindent\textit{Proof:} See Appendix~\ref{sec:appendix:proof of lemma 1}. \hfill $\blacksquare$

\begin{figure}[t!]
\centering
{\epsfig{figure=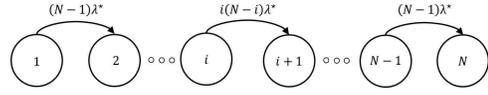,width=0.35\textwidth}}
\caption{Transition diagram of the Markov chain $\{I(t); t\ge 0\}$ when $K=1$.}
\label{fig:transition diagram when K=1}
\end{figure}

\smallskip\smallskip
\noindent\emph{Example I. (Homogeneous model, Single community model)} We start our analysis with the simplest case of $K=1$ (i.e., homogeneous model), and drop the group index in all notations for simplicity. In this case, we have $\bm{I}(t) = I(t)$ and $\mathcal{E}=\{1,\ldots,N\}$. Then, we can identify the temporal behavior of $I(t)$ as follows: first note that the process $\{I(t); t \ge 0\}$ is a \emph{counting process} in that it counts the number of events that have taken place during $(0,t]$. Hence, state transitions occur only to the adjacent state from $i\,(=1,\ldots,N-1)$ to $i+1$, and then eventually the system is absorbed to state~$N$. Thus, the state space $\mathcal{E}$ is decomposed into transient state space $\mathcal{E}^\star=\{1,\ldots,N-1\}$ and absorbing state space $\mathcal{E}^o=\{N\}$. Suppose that the system enters state $i\in\mathcal{E}^\star$ at time $t_0$. Let $X_i$ be the sojourn time of state~$i$. Note that the sojourn time is equivalent to the time to have one more infected node, which is the same as the minimum infection time from $i$ number of infected nodes to $N-i$ number of susceptible nodes, i.e.,
\begin{align*}
X_i = \min\big\{M_{a,b}^\text{eff}; \,a \in \mathcal{I}(t_0), b \in \mathcal{S}(t_0)\big\},
\end{align*}
where $\mathcal{I}(t)$ and $\mathcal{S}(t)$ denote index sets of infected nodes and susceptible nodes at time $t$, respectively. Since $M_{a,b}^{\text{eff}}\sim\text{Exp}(\lambda^\star)$ from~(\ref{eqn:exp meeting time:eff}) and is independent for all nodes, we have:
\begin{align*}
X_i\sim\text{Exp}(i(N-i)\lambda^\star).
\end{align*}
Therefore, the process $\{I(t); t\ge 0\}$ is a CTMC with transition diagram depicted in Fig.~\ref{fig:transition diagram when K=1}. From the transition diagram, we can easily obtain the matrix $\bm{F}$. For details, see Appendix~\ref{sec:appendix:fundamental matrix}.

\begin{figure}[t!]
\centering
{\epsfig{figure=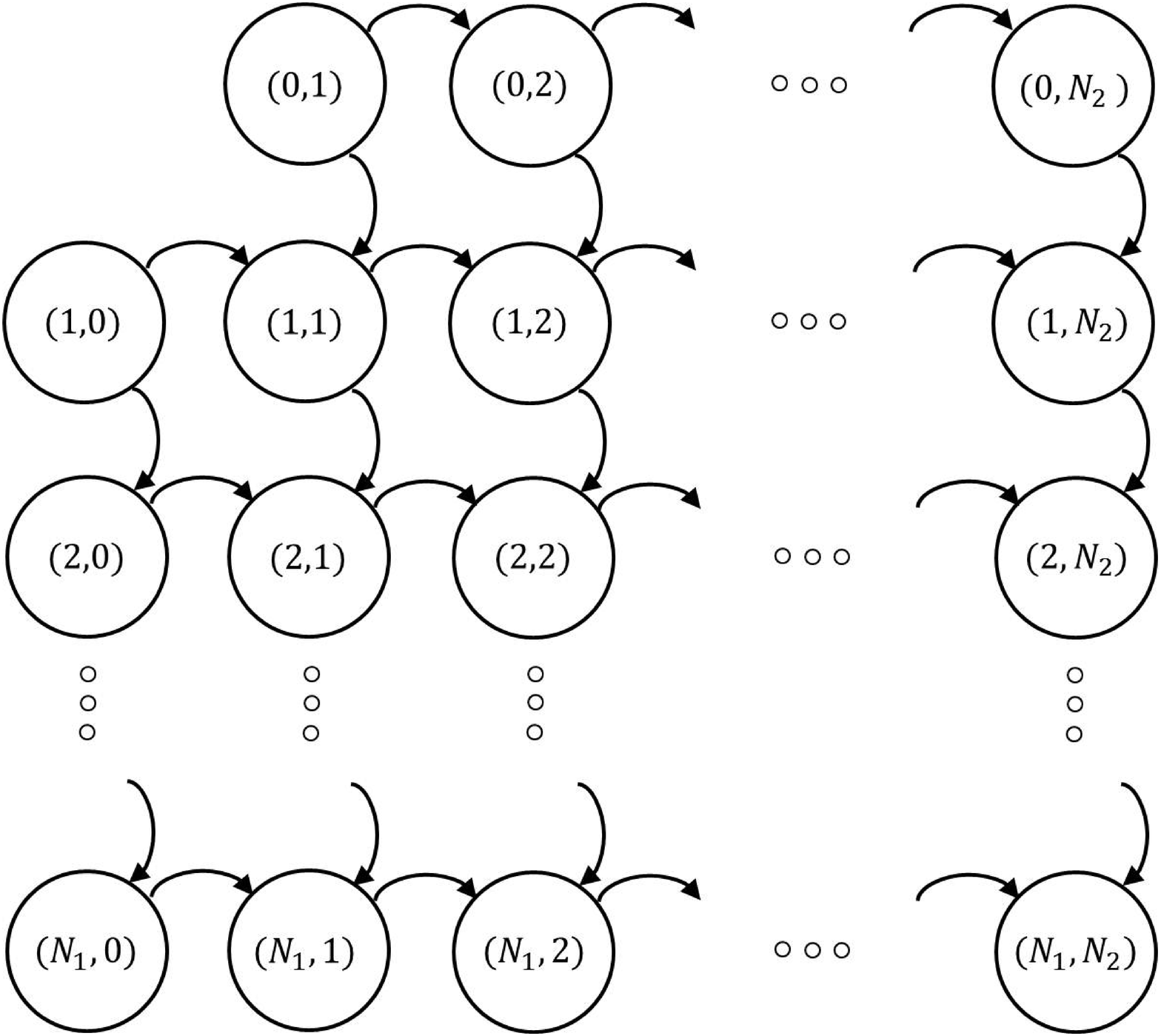,width=0.25\textwidth, height=0.20\textwidth}}
\caption{Transition diagram of the Markov chain $\{(I_1(t),I_2(t)); t\ge 0\}$ when $K=2$. The rate from $(i_1,i_2)$ to $(i_1+1,i_2)$ is $(N_1-i_1)\sum_{k=1}^{2}i_k\lambda_{k,1}^\star$, and the rate from $(i_1,i_2)$ to $(i_1,i_2+1)$ is $(N_2-i_2)\sum_{k=1}^{2}i_k\lambda_{k,2}^\star$.}
\label{fig:transition diagram when K=2}
\end{figure}

\smallskip\smallskip
\noindent\emph{Example II. (Double community model)} We next consider the case when $K=2$. In this case, if we set the state variable as the total number of infected nodes (i.e., $\sum_k I_k (t)$), then it becomes intractable to identify the statistics of sojourn time~$X_i$ of state $i$, unless we know how the overall infected nodes in the network are distributed to each group. For this reason, we set the vector $(I_1(t),I_2(t))$ as the state variable. Suppose that at time $t_0$, the system enters state $(i_1,i_2)$. Since the process $\{I_1(t)+I_2(t); t\ge 0\}$ is a counting process, the very next state transitions occur only to either $(i_1+1,i_2)$ or $(i_1,i_2+1)$, and then eventually the system is absorbed to state $(N_1,N_2)$. Hence, state space~$\mathcal{E}=\{(0,1),(1,0),(1,1), \ldots,(N_1,N_2)\}$ is decomposed into transient state space~$\mathcal{E}^\star=\{\bm{e}\in\mathcal{E}:|\bm{e}| < N_1+N_2\}$ and absorbing state space $\mathcal{E}^o=\{(N_1,N_2)\}$. For $(i_1,i_2)\in\mathcal{E}^\star$, let $X_{(i_1,i_2):i_1}$ and $X_{(i_1,i_2):i_2}$ be the time required to infect one additional node in groups~1 and~2, respectively. Then, by a similar reason as in Example~I, we have
\begin{align*}
X_{(i_1,i_2):i_1} = \min\big\{M_{a,b}^{\text{eff}}\,; a \in \mathcal{I}_1(t_0)\cup \mathcal{I}_2(t_0), b \in \mathcal{S}_1(t_0)\big\},
\end{align*}
where $\mathcal{I}_k(t)$ and $\mathcal{S}_k(t)\,(k=1,2)$ denote index sets of infected nodes and susceptible nodes in group~$k$ at time $t$, respectively. Thus, $X_{(i_1,i_2):i_1}$ follows an exponential distribution as in Example~I, but in this case the rate is given by  $i_1(N_1-i_1)\lambda_{1,1}^\star+i_2(N_1-i_1)\lambda_{2,1}^\star = (N_1-i_1)\sum_{k=1}^{2} i_k \lambda_{k,1}^\star$. Similarly, we have that $X_{(i_1,i_2):i_2}$ follows an exponential distribution as summarized below:
\begin{align}\label{eqn:K2 infect g1}
X_{(i_1,i_2):a} &\sim
\begin{cases}
\text{Exp}{\big((N_1-i_1)\sum_{k=1}^{2} i_k \lambda_{k,1}^\star\big)} & \text{if } a = i_1, \\
\text{Exp}{\big((N_2-i_2)\sum_{k=1}^{2} i_k \lambda_{k,2}^\star\big)} & \text{if } a = i_2.
\end{cases}
\end{align}
Note that the sojourn time $X_{(i_1,i_2)}$ of state $(i_1,i_2)$ is the minimum value between $X_{(i_1,i_2):i_1}$ and $X_{(i_1,i_2):i_2}$. Hence, from~(\ref{eqn:K2 infect g1}), $X_{(i_i,i_2)}$ follows an exponential distribution. Therefore, the process $\{(I_1(t),I_2(t));t\ge 0\}$ is a 2-dimensional CTMC with transition diagram depicted in Fig.~\ref{fig:transition diagram when K=2}. From the transition diagram, we can easily obtain the fundamental matrix $\bm{F}$. For details, see Appendix~\ref{sec:appendix:fundamental matrix}.

\begin{figure}[t!]
\centering
{\epsfig{figure=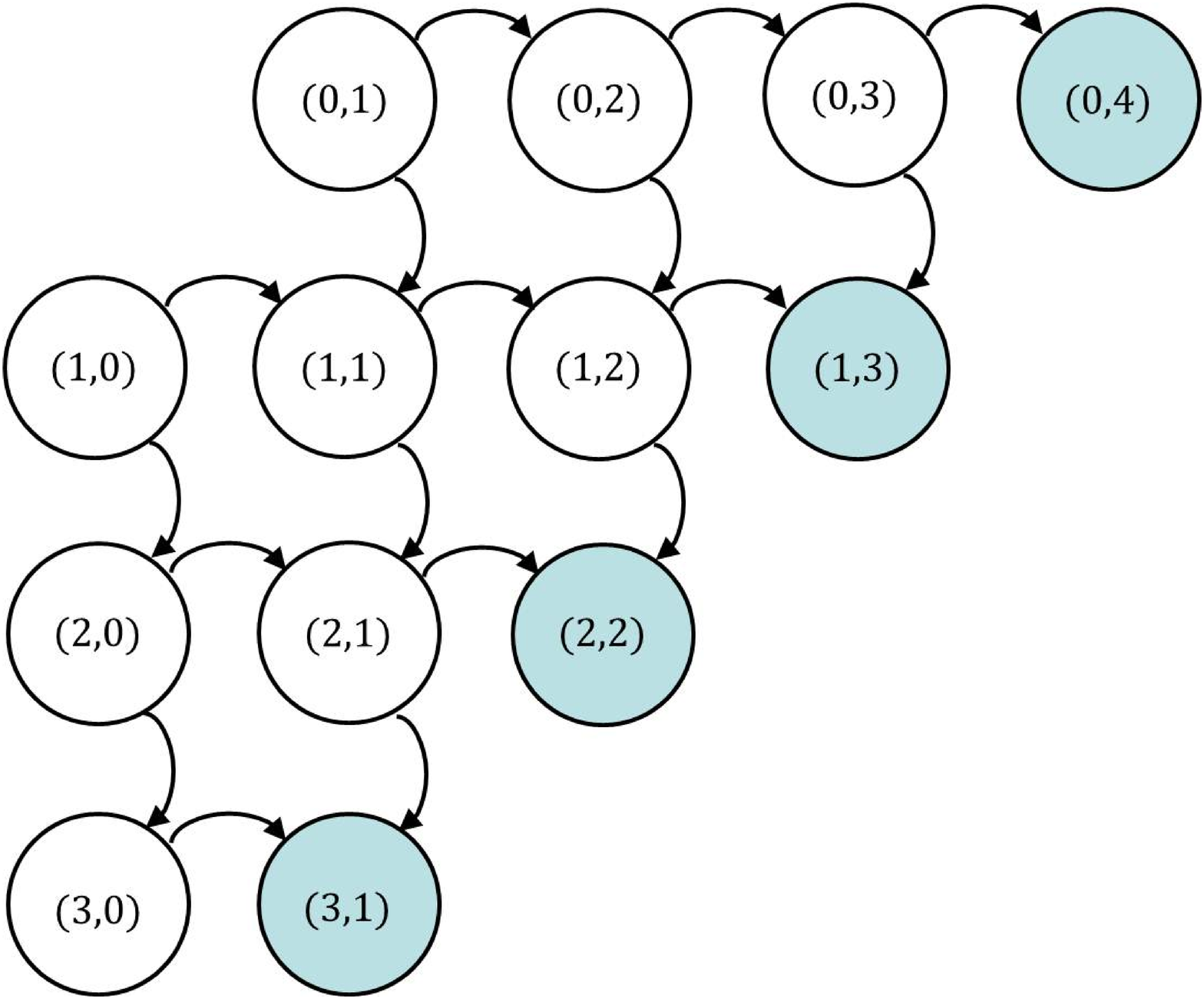,width=0.2\textwidth}}
\caption{An example of $\mathcal{E}_\alpha(=\mathcal{E}_\alpha^\star \cup \mathcal{E}_\alpha^o)$ for $(N_1,N_2) = (3,5)$, $\lceil \alpha N \rceil = 4$: shaded states form $\mathcal{E}_\alpha^o$, and the others form $\mathcal{E}_\alpha^\star$.}
\label{fig:truncated transition diagram when K=2}
\end{figure}

\smallskip
\noindent\emph{Step 2:} Using the results in Lemma~\ref{lemma:CTMC model}, we can derive the distribution of $T_\alpha$. We take two steps: (i) first, we truncate the state space $\mathcal{E}$ to $\mathcal{E}_\alpha \deq \{\bm{e}\in\mathcal{E}:|\bm{e}| \le \lceil\alpha N\rceil\}$, where $\lceil x\rceil $ denotes the smallest integer greater than or equal to $x$. (ii) Next, we split the truncated state space~$\mathcal{E}_\alpha$ into transient state space~$\mathcal{E}_\alpha^\star$ and absorbing state space~$\mathcal{E}_\alpha^o$ as:
\begin{align*}
\mathcal{E}_\alpha^\star &\deq \{\bm{e}\in\mathcal{E}_\alpha:|\bm{e}| < \lceil \alpha N \rceil \}, \\
\mathcal{E}_\alpha^o &\deq \{\bm{e}\in\mathcal{E}_\alpha:|\bm{e}| = \lceil \alpha N \rceil\}.
\end{align*}
On the state space~$\mathcal{E}_\alpha^\star\cup\,\mathcal{E}_\alpha^o$, we define a truncated process $\bm{I}_\alpha(t)$ from the process $\bm{I}(t)$ as follows: $\bm{I}_\alpha(t)$ evolves according to $\bm{I}(t)$ unless $\bm{I}(t) \in \mathcal{E}_\alpha^o$. When $\bm{I}(t)$ enters one of states in $\mathcal{E}_\alpha^o$, say $\bm{e}$, truncation happens and $\bm{I}_\alpha(t)$ is absorbed to the state~$\bm{e}$. Then, by Lemma~\ref{lemma:CTMC model} the process $\{\bm{I}_\alpha(t); t\ge 0\}$ is a $K$-dimensional CTMC with possibly multiple absorbing states in $\mathcal{E}_\alpha^o$. Moreover, by Definition~\ref{def:alpha completion time}, $T_\alpha$ is the time taken by the truncated Markov chain to be absorbed into $\mathcal{E}_\alpha^o$. An example of transition diagram is shown in Fig.~\ref{fig:truncated transition diagram when K=2}.

Similarly to (P2) in Lemma~\ref{lemma:CTMC model}, the infinitesimal generator~$\bm{Q}_\alpha$ of the process $\{\bm{I}_\alpha(t); t\ge 0\}$ is of the following form:
\begin{align*}
\bm{Q}_\alpha = \left[\begin{matrix}
\bm{F}_\alpha & \bm{F}^o_\alpha \\
\bm{0} & \bm{0}
\end{matrix}\right].
\end{align*}
Here, $\bm{F}_\alpha$ is a matrix representing transition rate from $\mathcal{E}^\star_\alpha$ to $\mathcal{E}^\star_\alpha$, and can be obtained from the fundamental matrix $\bm{F}$ of the original process $\{\bm{I}(t); t\ge 0\}$ by
\begin{align}\label{eqn:truncated F}
\bm{F}_\alpha = \bm{F}|_{\mathcal{E}^\star_\alpha\times\mathcal{E}^\star_\alpha}\, (=\bm{Q}|_{\mathcal{E}^\star_\alpha\times\mathcal{E}^\star_\alpha}).
\end{align}
Similarly, $\bm{F}^o_\alpha$ is a matrix representing transition rate from $\mathcal{E}^\star_\alpha$ to $\mathcal{E}^o_\alpha$, and is obtained by $\bm{F}_\alpha^o = \bm{Q}|_{\mathcal{E}^\star_\alpha\times\mathcal{E}^o_\alpha}.$ Therefore, the value~$\alpha$ determines where to truncate the matrix $\bm{F}$ or $\bm{Q}$ in Lemma~\ref{lemma:CTMC model} and how to redefine transient and absorbing state spaces. For $\alpha$ values satisfying $\lceil \alpha N \rceil = N$, we have $\mathcal{E}_\alpha^\star = \mathcal{E}^\star$ and $\mathcal{E}_\alpha^o= \mathcal{E}^o$, which gives $\bm{F}_\alpha = \bm{F}$ and $\bm{F}_\alpha^o = \bm{F}^o$.

Once we have the truncated fundamental matrix~$\bm{F}_\alpha$ from the original matrix~$\bm{F}$, we can obtain the distribution of $T_\alpha$ as in the following lemma.

\begin{lemma}[Distribution of $T_\alpha$]\label{lemma:distribution of completion time}
The cumulative distribution function (CDF) of the $\alpha$-completion time is given by
\begin{align*}
H_\alpha(t) \deq \pr\{T_\alpha \le t\} = 1- \bm{h}_\alpha \exp(\bm{F}_\alpha t)\bm{1},
\end{align*}
where $\bm{h}_\alpha \deq (P\{\bm{I}_\alpha(0)=\bm{e}\})_{\bm{e} \in\mathcal{E}_\alpha^\star}$ is a row vector denoting the initial distribution, $\bm{F}_\alpha$ is given in~(\ref{eqn:truncated F}), and $\bm{1}$ is a column vector of ones. In addition, it can be expressed as in the following form~\cite{aalen95}:
\begin{align*}
H_\alpha(t) = 1-\sum_{i}\exp(- |\rho_i| t) P_{i}(t),
\end{align*}
where $\rho_i$ denotes the $i$th eigenvalue of $\bm{F}_\alpha$ with multiplicity denoted by $m_i$, and $P_{i}(t)$ is a $(m_i-1)$th order polynomial function of~$t$. Since $\bm{F}_\alpha$ is an upper triangular matrix, the eigenvalues are from the distinct diagonal elements of the matrix, which are all real and negative.
\end{lemma}

\noindent\textit{Proof:} See Appendix~\ref{sec:appendix:proof of lemma 2}. \hfill $\blacksquare$

\smallskip

\smallskip
\noindent\emph{Step 3:} Based on Lemma~\ref{lemma:distribution of completion time}, we can derive formulas for our performance metrics, as shown in Lemma~\ref{lemma:formula for metrics}.

\begin{lemma}[Formulas for $G_{\alpha,\beta}$ and $R_{\alpha,\beta}$]\label{lemma:formula for metrics} The inverse function of the distribution function $H_\alpha(\cdot)$ in Lemma~\ref{lemma:distribution of completion time} exists and yields the $(\alpha,\beta)$-guaranteed time in~(\ref{eqn:def:guaranteed time}) as
\begin{align}\label{eqn:lemma:guaranteed time}
G_{\alpha,\beta} = H_\alpha^{-1}(\beta).
\end{align}
The fundamental matrix $\bm{F}_\alpha$ is invertible, and its inverse matrix gives the $n$th moment of $T_\alpha$ $(n=1,2,\ldots)$ as
\begin{align}\label{eqn:lemma:nth moment}
\E[(T_\alpha)^n] = n! \bm{h}_\alpha (-\bm{F}_\alpha)^{-n} \bm{1}.
\end{align}
Therefore, the ratio $R_{\alpha,\beta}$ is obtained from~(\ref{eqn:lemma:guaranteed time}) and~(\ref{eqn:lemma:nth moment}) by
\begin{align*}
R_{\alpha,\beta} = \frac{H_\alpha^{-1}(\beta)}{\bm{h}_\alpha (-\bm{F}_\alpha)^{-1} \bm{1}}.
\end{align*}
\end{lemma}

\noindent\textit{Proof:} See Appendix~\ref{sec:appendix:proof of lemma 3}. \hfill $\blacksquare$

\smallskip
Major applications leveraging $G_{\alpha, \beta} (= G_{\alpha, \beta}(\bm{\Lambda},\bm{s}))$ include the followings:
\begin{enumerate}
\item For distributing a firmware or a software update to smartphones (and tablets) through opportunistic contacts among nodes when cellular network carriers wish to avoid abusing network resources while guaranteeing the time to deliver the update with more than 99\% of confidence, $G_{\alpha,\beta}$ becomes significantly useful to determine the required number of seeds in the network. For instance, to guarantee delivery with probability$\beta$ for $\alpha$ fraction of nodes within time $T_{\text{bound}}$, the number of seeds $\bm{s}$ who directly get the update from the carriers can be determined from:
    \begin{equation*}
    \bm{s} = G^{-1}_{\alpha,\beta} (\bm{\Lambda}, T_{\text{bound}}).
    \end{equation*}
\item For an autonomous disaster broadcasting system, which purely leverages opportunistic contacts without relying on network infrastructures, the target level of infection rates $\bm{\Lambda}$, which achieves a desirable time bound $T_{\text{bound}}$, can be determined by:
    \begin{equation*}
    \bm{\Lambda} = G^{-1}_{\alpha,\beta} (T_{\text{bound}}, \bm{s})
    \end{equation*}
    for given $(\alpha, \beta)$. Based on this prediction, we can scale up or down the infection rates $\bm{\Lambda}$ among nodes by optimally controlling the communication ranges of mobile devices.
\item For a highly contagious disease emerged at a city, if medical facilities in the city have capacity for up to $\alpha$ portion of citizens who typically have $\bm{\Lambda}$ infection rates, the regional government can estimate the allowed time to execute emergency plans by referring to:
    \begin{equation*}
    T_{\text{bound}} = G_{\alpha, \beta} (\bm{\Lambda}, \bm{s}).
    \end{equation*}
\end{enumerate}

\section{Analytical Characteristics and Applications}\label{sec:characterization}
In this section, we present analytical characteristics derived from our framework, and provide how to utilize these characteristics in practical applications.

\subsection{Impact of the level of infection rates}
The behavior of information spread is determined by various spreading factors. Using our framework, we first answer the question on how the level of infection rates $\lambda^{\text{eff}}_{a,b}$ affect the distribution of $\alpha$-completion time.

\begin{theorem} [Impact of the level of infection rates]\label{thm:impact of lambda level}
Suppose that the infection rate $\lambda^\text{eff}_{a,b}$ is scaled by $\gamma\,(>0)$ times for all $a,b$. Let $\hat{T}_\alpha$, $\hat{G}_{\alpha,\beta}$, and $\hat{R}_{\alpha,\beta}$ be the correspondences of $T_\alpha$, $G_{\alpha,\beta}$, and $R_{\alpha,\beta}$ after the scale, respectively. Then, for any $\alpha\in(0,1]$, we have
\begin{align}\label{eqn:thm:lambda level}
\hat{T}_\alpha \ed \gamma^{-1}T_\alpha,
\end{align}
where $\ed$ denotes ``equal in distribution." The relationship in~(\ref{eqn:thm:lambda level}) yields for any $\alpha\in(0,1]$ and $\beta\in(0,1)$ the followings:
\begin{align*}
\hat{G}_{\alpha,\beta} &= \gamma^{-1}G_{\alpha,\beta}, \\
E[(\hat{T}_\alpha)^n] &= \gamma^{-n} E[(T_\alpha)^n].
\end{align*}
Hence, $\hat{R}_{\alpha, \beta}$ becomes $\hat{R}_{\alpha, \beta} = R_{\alpha,\beta}$.
\end{theorem}

\noindent\textit{Proof:} See Appendix~\ref{sec:appendix:proof of thm 1}. \hfill $\blacksquare$

\smallskip The result in Theorem~\ref{thm:impact of lambda level} shows that the spread becomes faster \emph{proportionally} to the level of infection rates in \emph{distribution sense}. Similarly, we show that the average $\mathcal{M}(t) \deq \sum_k \E[I_k (t)]$ and its time derivative $\mathcal{D}(t) \deq \frac{d}{dt}\mathcal{M}(t)$ scale respectively as $\hat{\mathcal{M}}(t)=\mathcal{M}(\gamma t)$ and $\hat{\mathcal{D}}(t)=\gamma\mathcal{D}(\gamma t)$ for all $t\ge0$. The detailed proof is given in Appendix~\ref{sec:appendix:proof of avg scaling}.

\subsection{Impact of population size}
We next characterize the impact of population size on information spread. In our epidemic model, each non-seed node can be considered as a workload to finish. However, once the node becomes infected, it works in a similar manner as the seed and is involved in spreading the information. Hence, it is not straightforward whether the population size accelerates or slows down the speed of information spread. Our framework gives the answer, as shown in Theorem~\ref{thm:impact of population size}.

\begin{theorem}[Impact of population size]\label{thm:impact of population size}
Suppose $\alpha=1$ (i.e., spread completion), $K=1$ (i.e., homogeneous model), and $s_1 = 1$ (i.e., one seed). As the population size~$N$ increases,
we have
\begin{itemize}
\item $G_{\alpha,\beta}$ is strictly decreasing for sufficiently large~$\beta$.
\item $\E[T_\alpha]$ is strictly decreasing.
\end{itemize}
In addition, it scales respectively as
\begin{itemize}
\item $G_{\alpha,\beta} = \Theta\big((\lambda^\star)^{-1} N^{-1}(\log N-\log(\log\frac{1}{\beta}))\big).$
\item $\E[T_\alpha] = \Theta\big((\lambda^\star)^{-1} N^{-1}\log N\big).$
\end{itemize}
Hence, $R_{\alpha,\beta}$ scales as $\Theta(1)$.
\end{theorem}

\noindent\textit{Proof:} See Appendix~\ref{sec:appendix:proof of thm 2}. \hfill $\blacksquare$

\smallskip The results in Theorem~\ref{thm:impact of population size} indicate that adding a node in the system accelerates the information spread when per-pair infection rates are unchanged.

\begin{remark}\label{rmk} To assist understanding of Theorem~\ref{thm:impact of population size}, we consider a non-cooperative spread model, where seed chosen at the beginning only spreads the information.\footnote{In epidemiological term, this non-cooperative model is classified as a SIR model with zero recovery time from infection.} As the population size~$N$ increases, we have
\begin{itemize}
\item $G_{\alpha,\beta}$ is strictly increasing for sufficiently large~$\beta$.
\item $\E[T_\alpha]$ is strictly increasing.
\end{itemize}
In addition, it scales respectively as
\begin{itemize}
\item $G_{\alpha,\beta} = \Theta\big((\lambda^\star )^{-1}(\log N-\log(\log\frac{1}{\beta}))\big).$
\item $\E[T_\alpha] = \Theta\big((\lambda^\star )^{-1}\log N\big).$
\end{itemize}
Hence, $R_{\alpha,\beta}$ scales as $\Theta(1)$.
\end{remark}
More properties of our model (namely, cooperative model) and the non-cooperative are compared in the following table:
\vspace{-0.2cm}
\begin{center}{\footnotesize
\begin{tabular}{|c|c|c|} \hline
    & Cooperative model & Non-cooperative model  \\ \hline \hline
Variance 	&  Strictly decrease with $N$ & Strictly increase with $N$ and\\
of $T_\alpha$	& and scale as $\Theta((\lambda^\star N)^{-2})$ & converge to $(\lambda^\star)^{-2}\zeta(2)$ \\  \hline
Skewness 	& Strictly decrease with $N$ & Strictly increase with $N$ and\\
 of $T_\alpha$	& and scale as $\Theta((\lambda^\star N)^{-3})$ & converge to $(\lambda^\star)^{-3}\zeta(3)$ \\ \hline
$\E[(T_\alpha)^n]$	& $\E[(T_\alpha)^n] <\infty$ & $\E[(T_\alpha)^n] = \infty$ \\ \hline
\end{tabular}}
\end{center}

In the table, $\zeta(c)\deq \sum_{n=1}^\infty n^{-c}$ denotes the Riemann zeta function. The proof for the results in Remark~\ref{rmk} is given in Appendix~\ref{sec:appendix:proof of rmk}. Our analysis showing that $G_{\alpha, \beta}$ behaves differently for the scaling of $N$ and $\lambda^\star$ tells that resource allocation for information spread should be carefully designed based on the willingness of cooperation in a spread process (i.e., infectivity in a spread process).

\subsection{Impact of multiple community}
The impact of heterogeneity in information or virus spreading has been less explored. Using our CTMC-based framework, we analyze and understand the temporal spread behavior under a heterogeneous network with multiple groups compared with a homogeneous network. In particular, we focus on answering ``Does heterogeneity persistently expedite the spreading or not?", ``Is there an optimal heterogeneity level for information spread?", and ``Is there an upper or a lower bound on the gain from the heterogeneity over homogeneity?".

In this subsection, we provide the answers to these questions by studying dual community model ($K=2$). Note that our framework can be easily extended to study the cases when $K \geq 3$. In order to focus on heterogeneity arising from multiple community, we make assumptions as follows: (i) two groups are of the same size $N_1 =  N_2 (=N/2)$. (ii) The inter-group infection rates are the same for both directions, i.e., $\lambda_{1,2}^\star=\lambda_{2,1}^\star$ (iii) There is one seed. Without loss of generality, the seed is chosen arbitrarily from group~1.

\begin{figure}[t!]
  \centering{
  \subfigure{\epsfig{figure=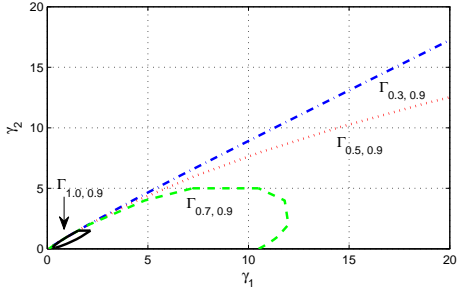,width=0.38\textwidth}}}
  \caption{Comparison of $(\alpha,\beta)$-guaranteed time $G_{\alpha,\beta}$ with the homogeneous model for $\beta = 0.9$ and $\alpha=0.3,0.5,0.7,1.0$: if $(\gamma_1,\gamma_2) \in \Gamma_{\alpha,\beta}$, then heterogeneity in multiple community accelerates the information spread (i.e., reduces the guaranteed time $G_{\alpha,\beta}$). If $(\gamma_1,\gamma_2) \notin \Gamma_{\alpha,\beta}$, then heterogeneity slows down the information spread.}
  \label{fig:scenarioII:guaranteed time}
\end{figure}

Let $\gamma_1 \deq \lambda_{1,1}^\star/\lambda_{1,2}^\star$ and $\gamma_2 \deq \lambda_{2,2}^\star/\lambda_{1,2}^\star$. The values of $\gamma_1$ and $\gamma_2$ control the intra-group infection rates, and are chosen freely in the range $0 \le \gamma_1, \gamma_2 <\infty$. Note that $(\gamma_1, \gamma_2)=(1,1)$ reduces to the homogeneous model and larger deviation from $(1,1)$ induces more heterogeneity. For a fair comparison with a homogeneous model of size $N$ and infection rate $\lambda^\star$, we use the following constraint that represents the same average infection rate:
\begin{align}\label{eqn:constraint}
\lambda^\star = \frac{\sum_{a}\sum_{b \neq a}\lambda_{g(a),g(b)}^\star}{N(N-1)}.
\end{align}
With the help of Theorems~\ref{thm:impact of lambda level} and~\ref{thm:impact of population size} showing the scaling of $\lambda^\star$ and $N$, we can characterize and generalize the impact of heterogeneity by only observing a specific setting of $(\lambda^\star, N)$. For simplicity, we choose $(1,40)$. We then vary $(\gamma_1,\gamma_2)$ in the range $0 \le \gamma_1, \gamma_2 \le 20$. From Lemma~\ref{lemma:formula for metrics}, we obtain the $(\alpha,\beta)$-guaranteed time~$G_{\alpha,\beta}$ and compare it with the homogeneous counterpart. Fig.~\ref{fig:scenarioII:guaranteed time} shows the result. In the figure, $\Gamma_{\alpha,\beta}$ is the region such that if $(\gamma_1,\gamma_2)\in\Gamma_{\alpha,\beta}$, then heterogeneity yields reduced guaranteed time $G_{\alpha,\beta}$, compared with the homogeneous model, and vice versa. Hence, the region $\Gamma_{\alpha,\beta}$ can be interpreted as the area where heterogeneity accelerates the information spread. From the figure, we can observe the followings: (i) as $\alpha$ increases, the region~$\Gamma_{\alpha,\beta}$ shrinks. Hence, for a fixed $(\gamma_1,\gamma_2)$, there exists a threshold $\alpha_\text{th}$ such that $(\gamma_1,\gamma_2) \in \Gamma_{\alpha,\beta}$ if $\alpha \le \alpha_\text{th}$ and $(\gamma_1,\gamma_2) \notin \Gamma_{\alpha,\beta}$ if $\alpha > \alpha_\text{th}$. In addition, the threshold decreases as $(\gamma_1,\gamma_2)$ deviates from (1,1). This implies that heterogeneity accelerates the spread at beginning phase (i.e., $\alpha \leq \alpha_{\text{th}}$) while slowing down the spread at ending phase (i.e., $\alpha > \alpha_{\text{th}}$), and the time portion of being accelerated shrinks with more heterogeneity. (ii) For any $\alpha\in\{0.3,0.5,0.7,1.0\}$, there is a non-empty region~$\bigcap_{\alpha} \Gamma_{\alpha,\beta}$, where \emph{heterogeneity always accelerates the information spread} (i.e., $\alpha_{\text{th}}=1$ for all $\alpha$). (iii) In the region $\{(\gamma_1,\gamma_2): \gamma_1 < \gamma_2\}$, heterogeneity always slows down the information spread. That is, if the seed is chosen from a less infective group, then heterogeneity never accelerates the information spread.

\begin{figure}[t!]
  \centering{
  \subfigure{\epsfig{figure=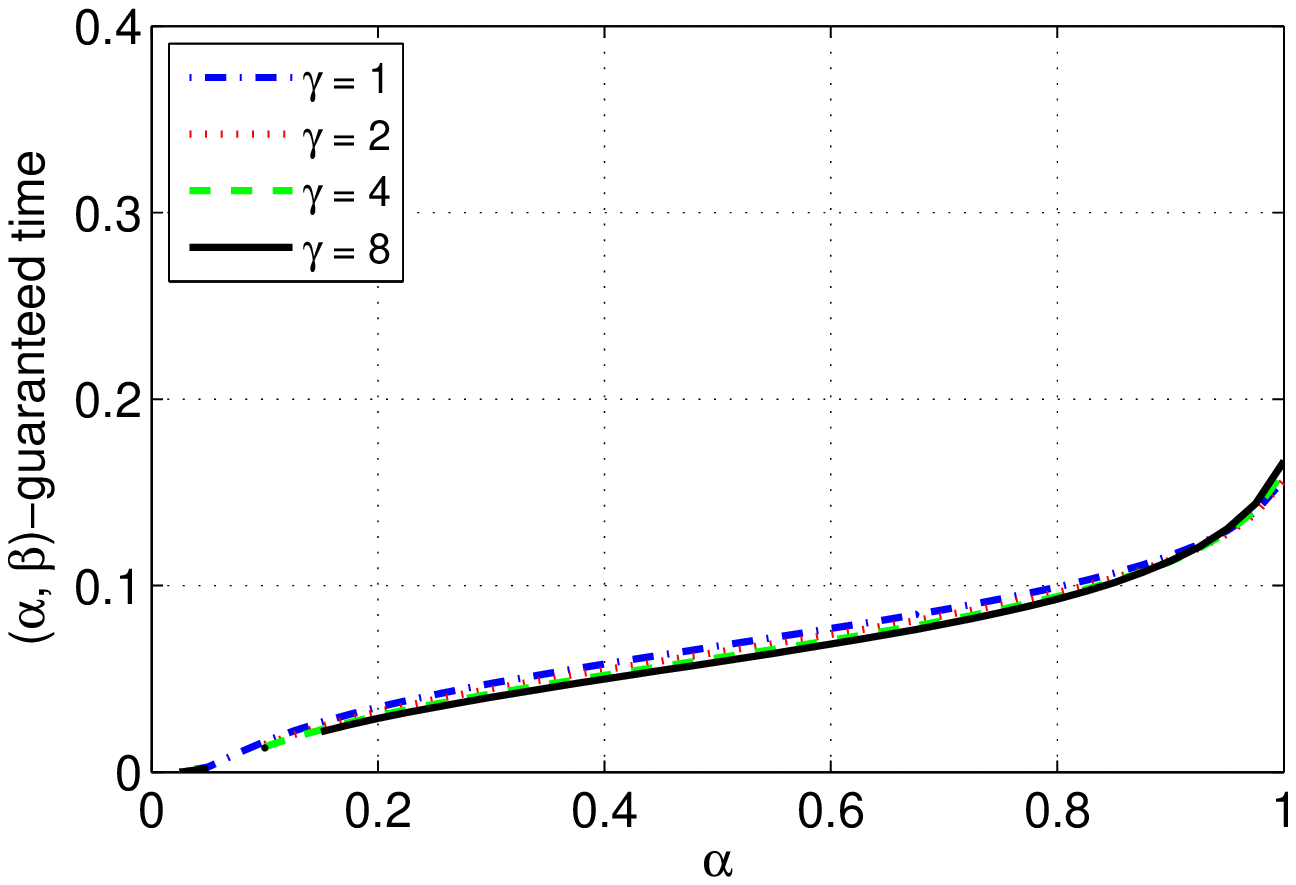,width=0.24\textwidth}}
  \subfigure{\epsfig{figure=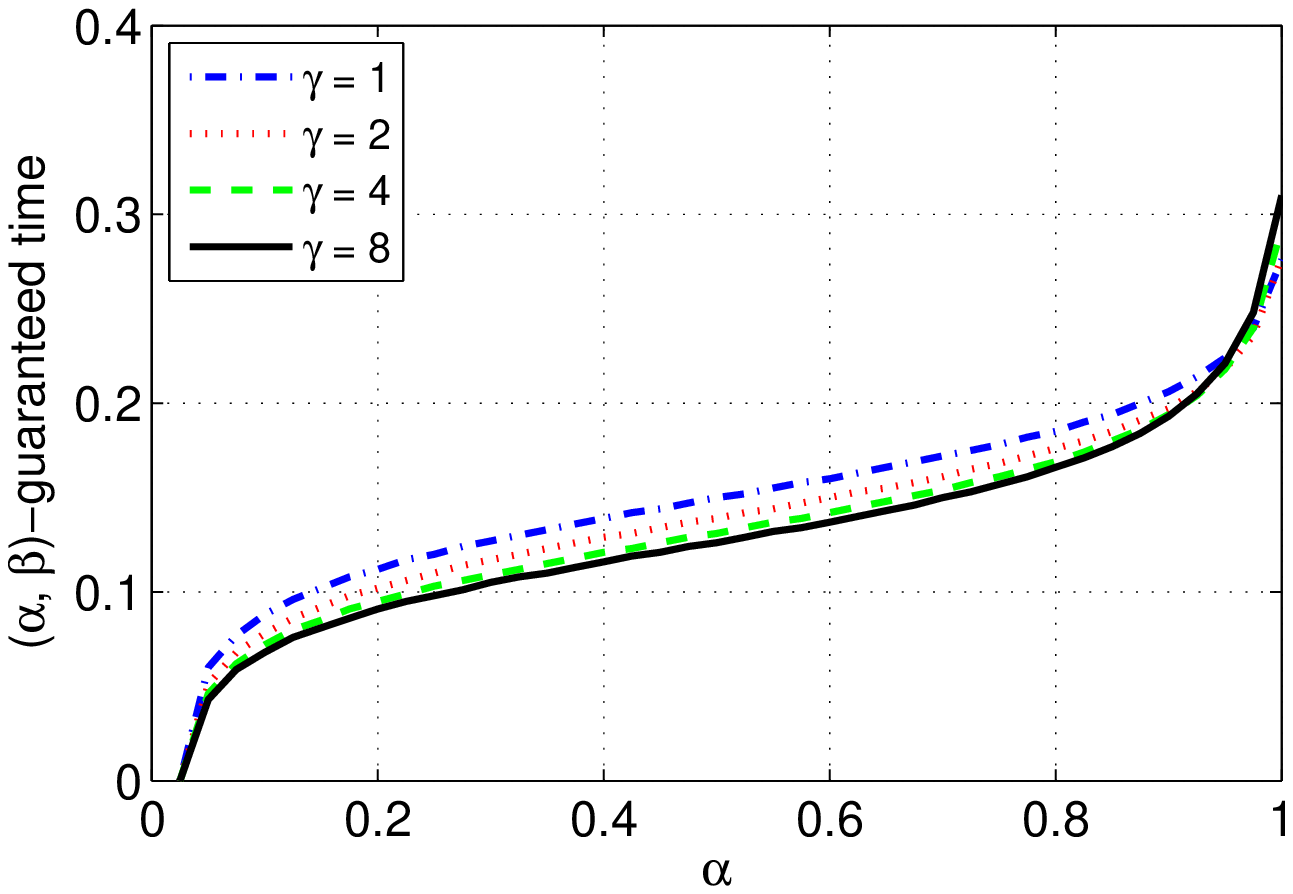,width=0.24\textwidth}}}
  \caption{$(\alpha,\beta)$-guaranteed time $G_{\alpha,\beta}$ for $\beta = 0.1$ (left) and $\beta = 0.9$ (right).}
  \label{fig:scenarioI:guaranteed time}
\end{figure}

As a special case, we consider a system where the inter-group infection rate is determined from intra-group infection rates by $\lambda_{1,2}^\star = (\lambda_{1,1}^\star+\lambda_{2,2}^\star)/2$, and the seed is chosen from more infective group. Let $\gamma \deq \max\{\lambda_{1,1}^\star,\lambda_{2,2}^\star\} /\min\{\lambda_{1,1}^\star,\lambda_{2,2}^\star\}.$ For fixed $(\lambda^\star,N)=(1,40)$ as above, we vary $\gamma$ as $\gamma =1,2,4,8$, and show the $(\alpha,\beta)$-guaranteed time $G_{\alpha,\beta}$ in Fig.~\ref{fig:scenarioI:guaranteed time}. From the figure, we confirm that heterogeneity indeed accelerates the spread for smaller penetration (i.e., for low $\alpha$) but slows down it for higher penetration. This observation is proved in Theorem~\ref{thm:impact of multiple community}.

\begin{theorem}[Impact of multiple community]\label{thm:impact of multiple community} Let
\begin{align*}
D_{\alpha}(\gamma)\deq -\lim_{t\to\infty}\frac{\log\pr\{T_{\alpha}(\gamma)>t\}}{t},
\end{align*}
where $T_\alpha(\gamma)$ denotes the $\alpha$-completion time when $\gamma$ is used. Then, $D_{\alpha}(\gamma)$ exists and satisfies the followings:
\begin{itemize}
\item If $\alpha\le 1-\frac{2}{N}$, then $\frac{\text{d}}{\text{d}\gamma}D_{\alpha}(\gamma) >0$ for all $\gamma\ge 1$.
\item If $\alpha= 1$, then $\frac{\text{d}}{\text{d}\gamma}D_{\alpha}(\gamma) <0$ for all $\gamma\ge 1$.
\item If $1-\frac{2}{N} < \alpha < 1$, then $\frac{\text{d}}{\text{d}\gamma}D_{\alpha}(\gamma) >0$ for $\gamma< \frac{5N-16}{N-4}$ and $\frac{\text{d}}{\text{d}\gamma}D_{\alpha}(\gamma) <0$ for $\gamma > \frac{5N-16}{N-4}$.
\end{itemize}
\end{theorem}

\noindent\textit{Proof:} See Appendix~\ref{sec:appendix:proof of thm 3}. \hfill $\blacksquare$

\subsection{Contribution of each node to the information spread}
In this section, we provide a method for quantifying the contribution of each individual node to the information spread. The quantification can be useful, e.g., for cellular carriers in incentivizing a node who contributes to alleviate data deluge in cellular networks by distributing packets through opportunistic contacts among nodes. Such an evaluation tool is of importance especially when nodes have heterogeneous attributes in spreading the information. Let $C_i$ denote the degree of contribution of node~$i$ to the spread. In this work, we evaluate $C_i$ by using the concept of the \emph{Shapely value}~\cite{shapley53}, which is known as a good metric measuring the surplus (or the contribution) of a node in the cooperative game theory:
\begin{align}\label{eqn:contribution}
C_i \deq \frac{G_{\alpha,\beta}(\mathcal{N}\setminus\{i\})}{G_{\alpha (N-1)/N ,\beta}(\mathcal{N})},
\end{align}
where $\mathcal{N}\deq\{1,\ldots,N\}$ is the index set of nodes, and for an index set $\mathcal{A}$, $G_{\alpha,\beta}(\mathcal{A})$ is the $(\alpha,\beta)$-guaranteed time for the network consisting of nodes $i\in\mathcal{A}$. Hence, the numerator and the denominator in~(\ref{eqn:contribution}) denote the guaranteed times in the network \emph{without} and \emph{with} the node~$i$, respectively, and consequently a node $i$ with high contribution to the information spread has a large $C_i$ value. Due to page limitation, we omit detailed application of the metric $C_i$ and its analysis.

\subsection{Applications}
How to optimally distribute given resources to nodes in a network to minimize the time for spreading of information to the network is of an important research question. Our results presented in this section provide initial understanding to this question. Theorem~\ref{thm:impact of multiple community} proves that when the number of nodes $N$ increases, heterogeneity in $\bm{\Lambda}$ expedites the spread of information for most of the time except some time duration at the end of spread, where the duration converges to zero as $N$ goes to infinity. It is important to point out that our understanding implies the existence of a small region of $\bm{\Lambda}$ with heterogeneous contact rates, which always make the spread faster than a network with homogeneous $\bm{\Lambda}$. By applying these two observations to designing a network, we have the following applications:

\begin{enumerate}
\item For a network delivering information to a community using vehicles or message ferries (e.g., DakNet\cite{pentland04}, DieselNet\cite{burgess06}, and ZebraNet\cite{juang02}) of which total amount of fuel is given, the amount of fuel distributed to each vehicle can be asymmetric to guarantee faster spread of information all the time compared to symmetric distribution.
\item When the number of nodes in a network is extremely large (e.g., users in facebook), advertising a product to the network can be expedited by providing incentives to users to forward information to others in a highly skewed manner. Our results support that evenly distributed incentives to the entire population would lead to much slower spreading compared to unfair incentives. This tells that the same speed of spread can be achieved by only providing a smaller amount of total incentive to the network when incentives are optimally distributed with the understanding of skewness.
\end{enumerate}

\section{Simulation Study}\label{sec:numerical}
\begin{figure*}[t]
\centering
\subfigure[Number of contacts for each vehicle] {\epsfig{figure=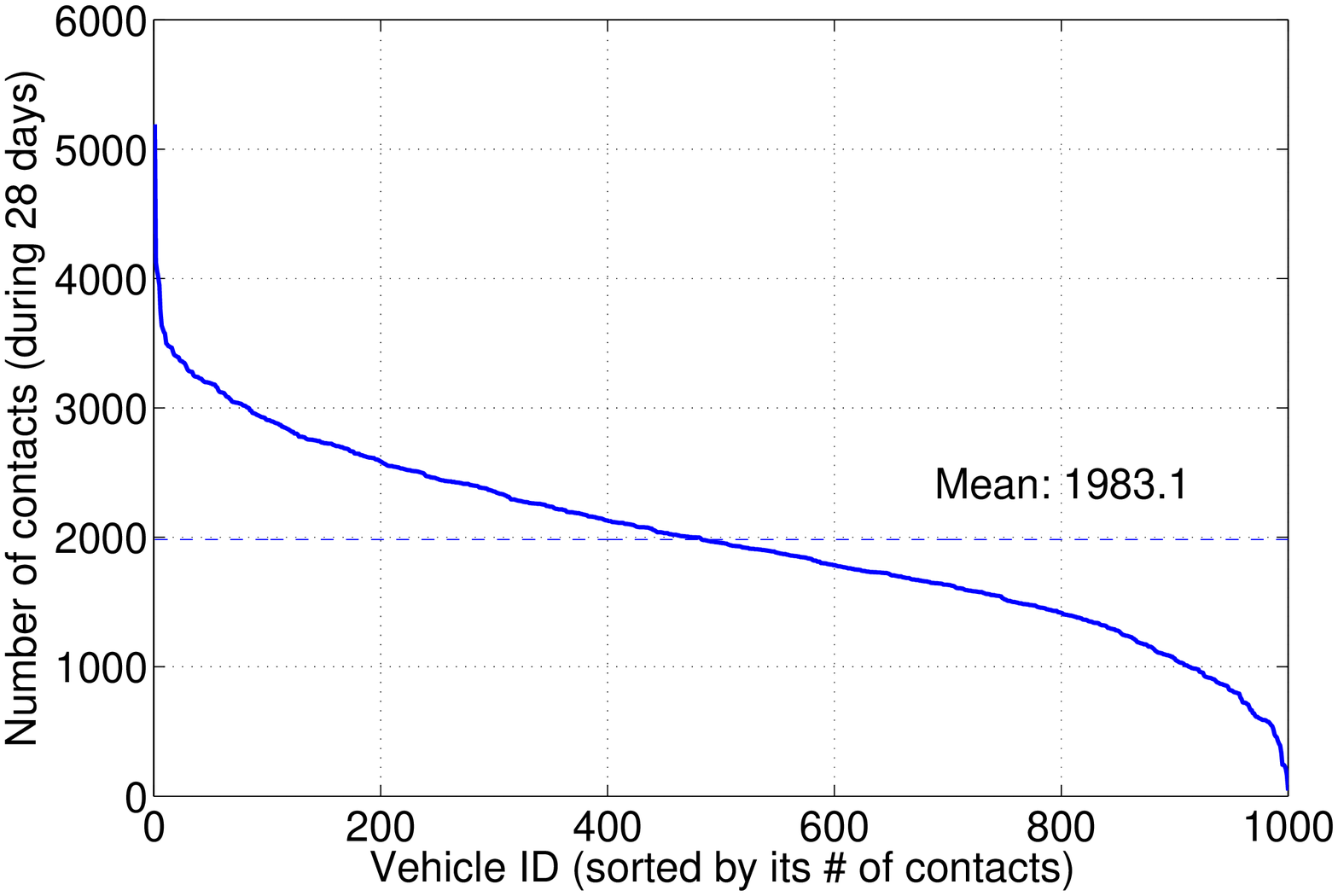,width=0.26\textwidth}}
\hspace{0.3cm}
\subfigure[Average number of neighbors] {\epsfig{figure=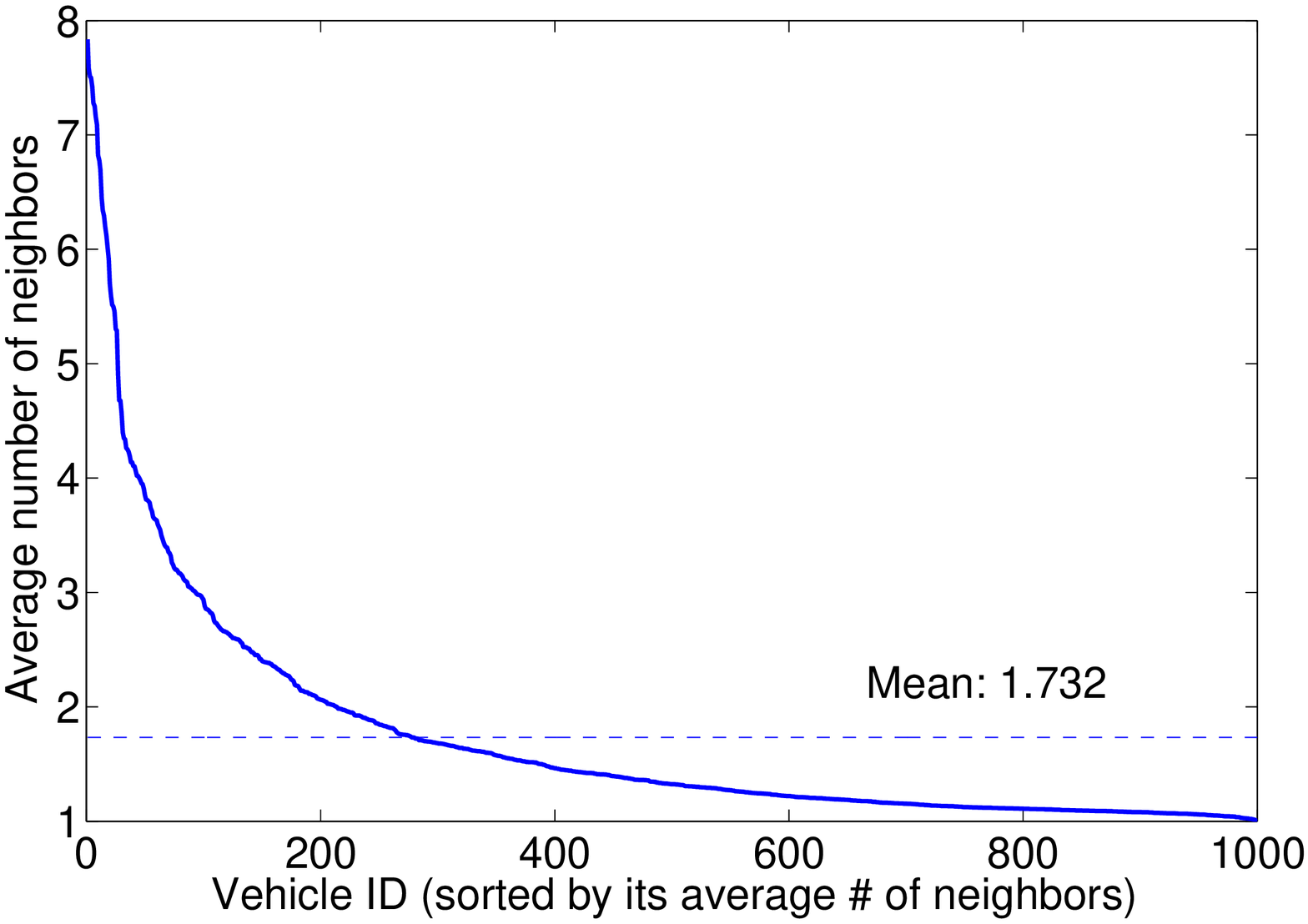, width=0.26\textwidth}}
\hspace{0.3cm}
\subfigure[CDF of aggregated contact durations] {\epsfig{figure=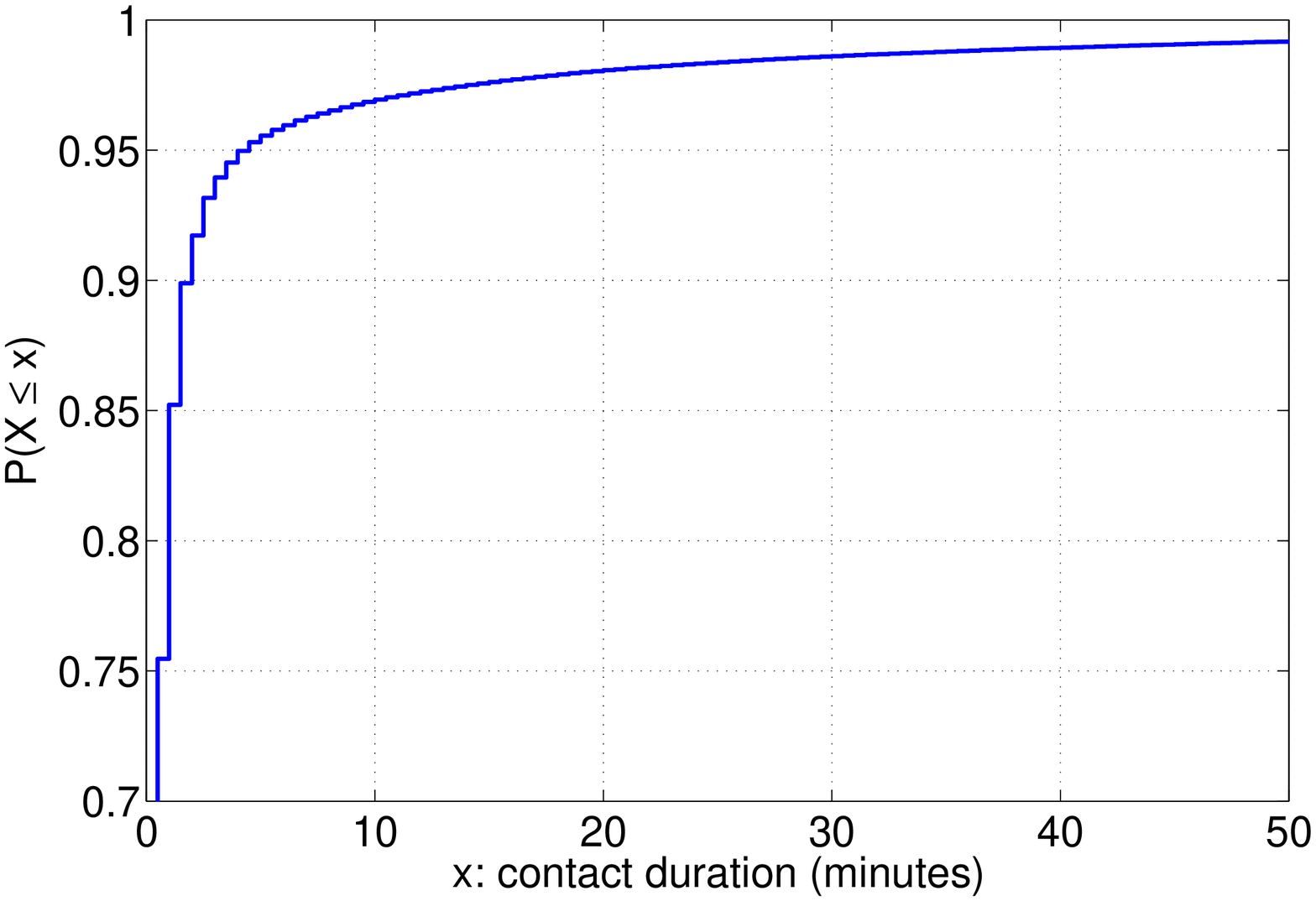, width=0.26\textwidth}}
\caption{(a) Number of contacts of a vehicle with all other vehicles during 28 days. (b) Average number of neighbors when a node is in a contact with another node. (c) CDF of aggregated contact durations between all taxi pairs.}
\label{fig:shanghai}
\end{figure*}

\begin{figure*}[t]
\centering
\subfigure[$(\alpha, \beta)$-guaranteed time with 1 seed] {\epsfig{figure=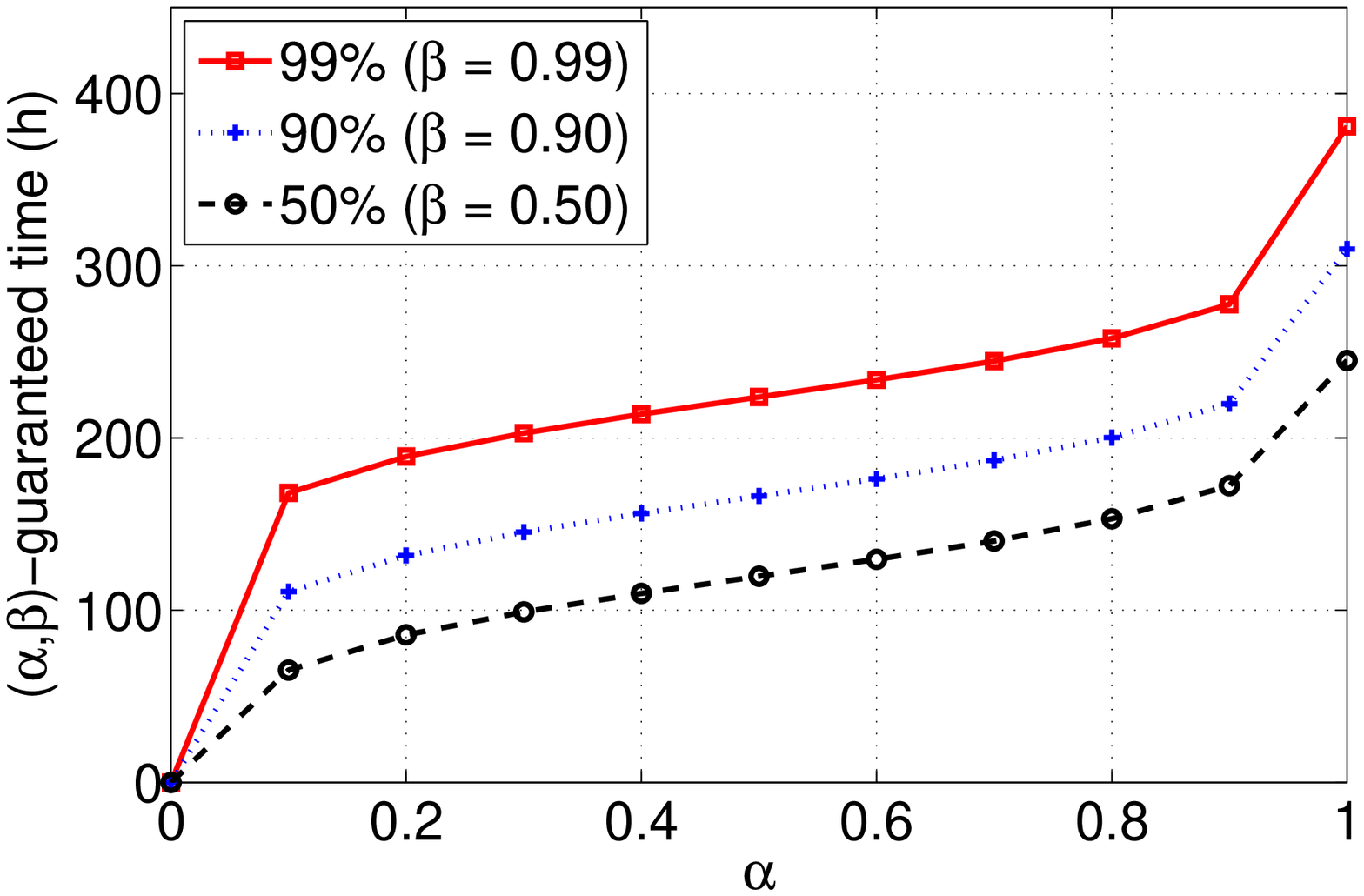,width=0.26\textwidth}} \hspace{0.3cm}
\subfigure[$(\alpha, \beta)$-guaranteed time with 10 seeds] {\epsfig{figure=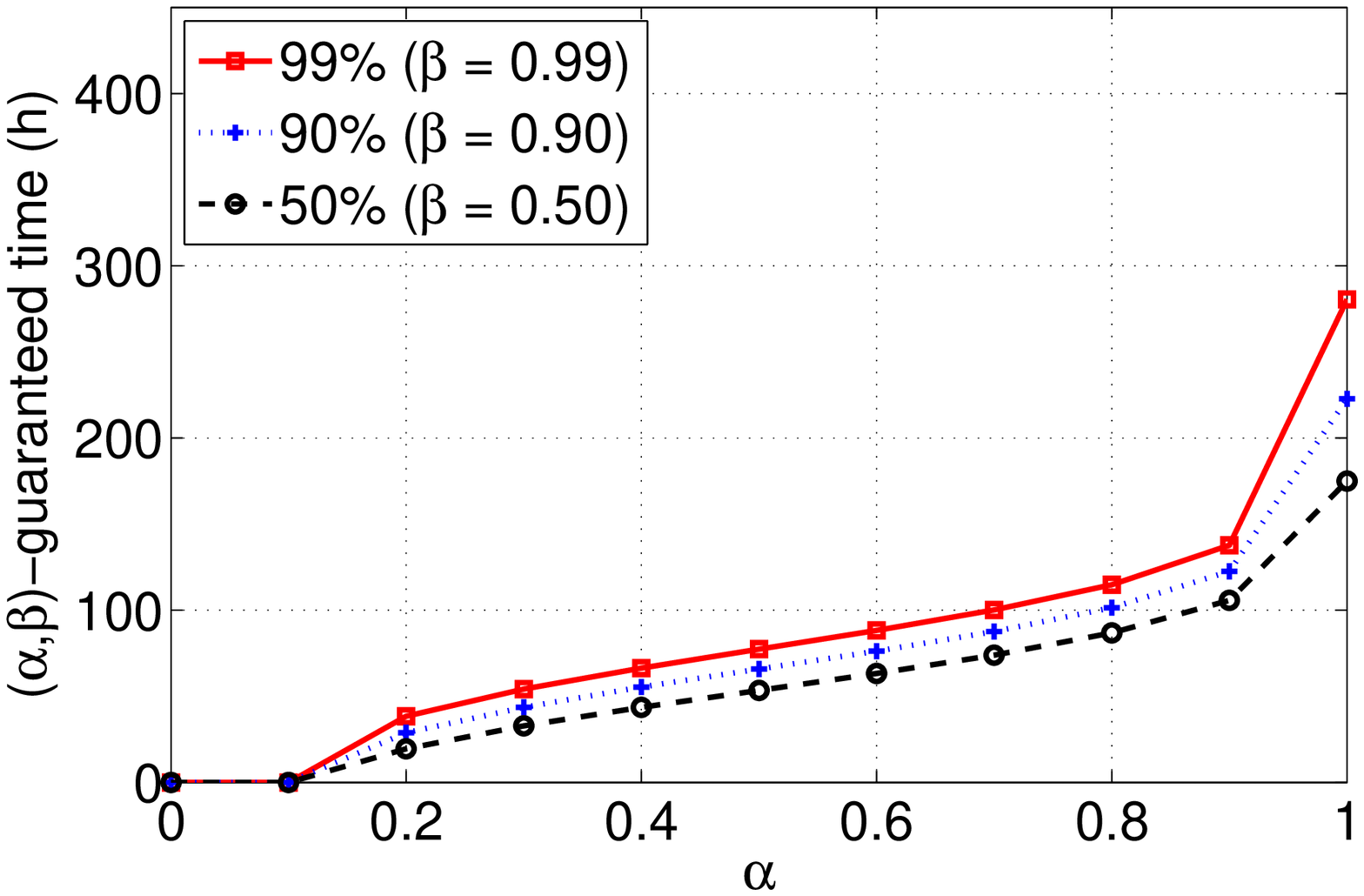, width=0.26\textwidth}}
\hspace{0.3cm}
\subfigure[$(\alpha, \beta)$-guaranteed time with 20 seeds] {\epsfig{figure=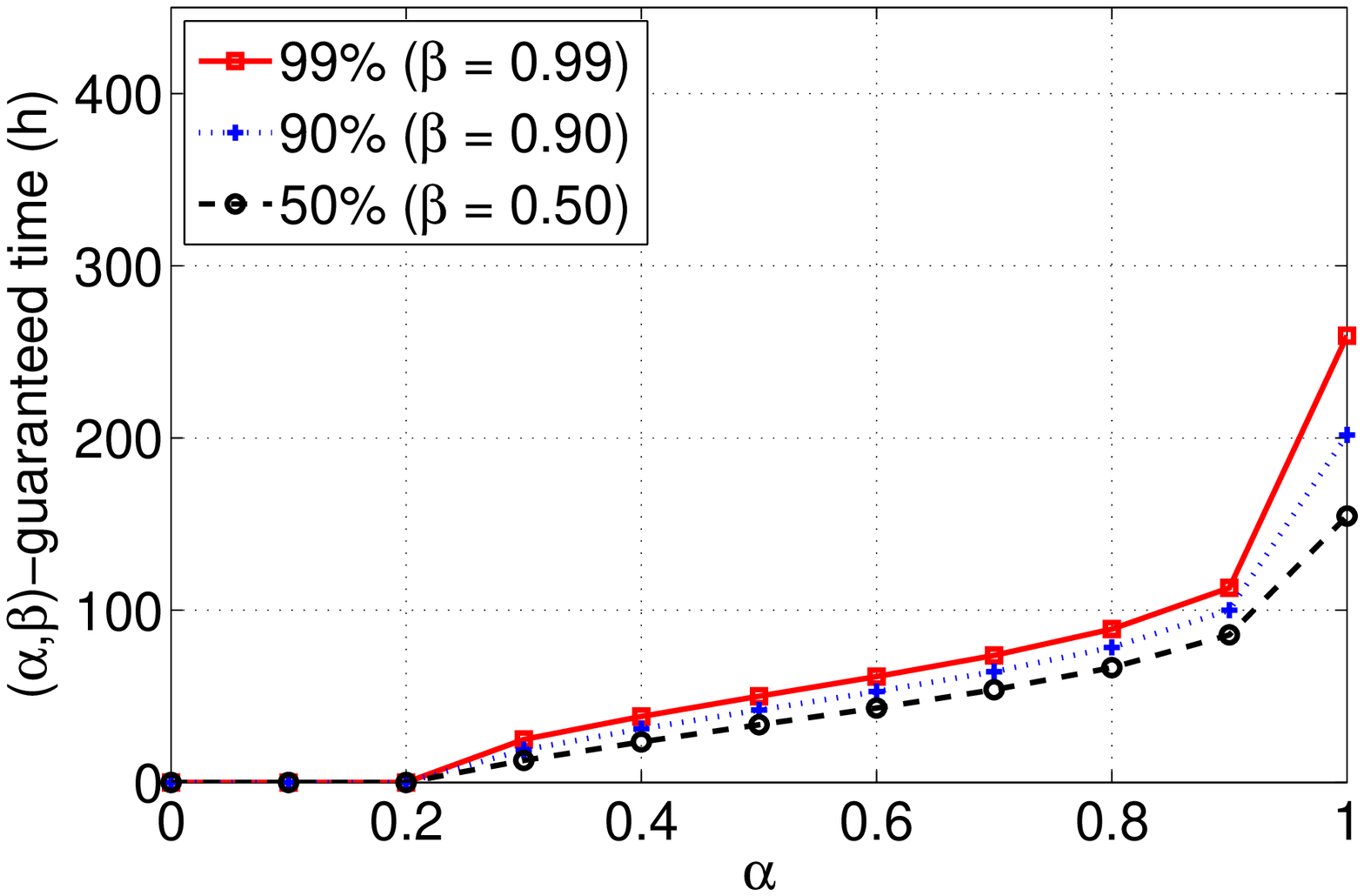, width=0.26\textwidth}}\\
\subfigure[$(\alpha, \beta)$-guaranteed time with 1 seed] {\epsfig{figure=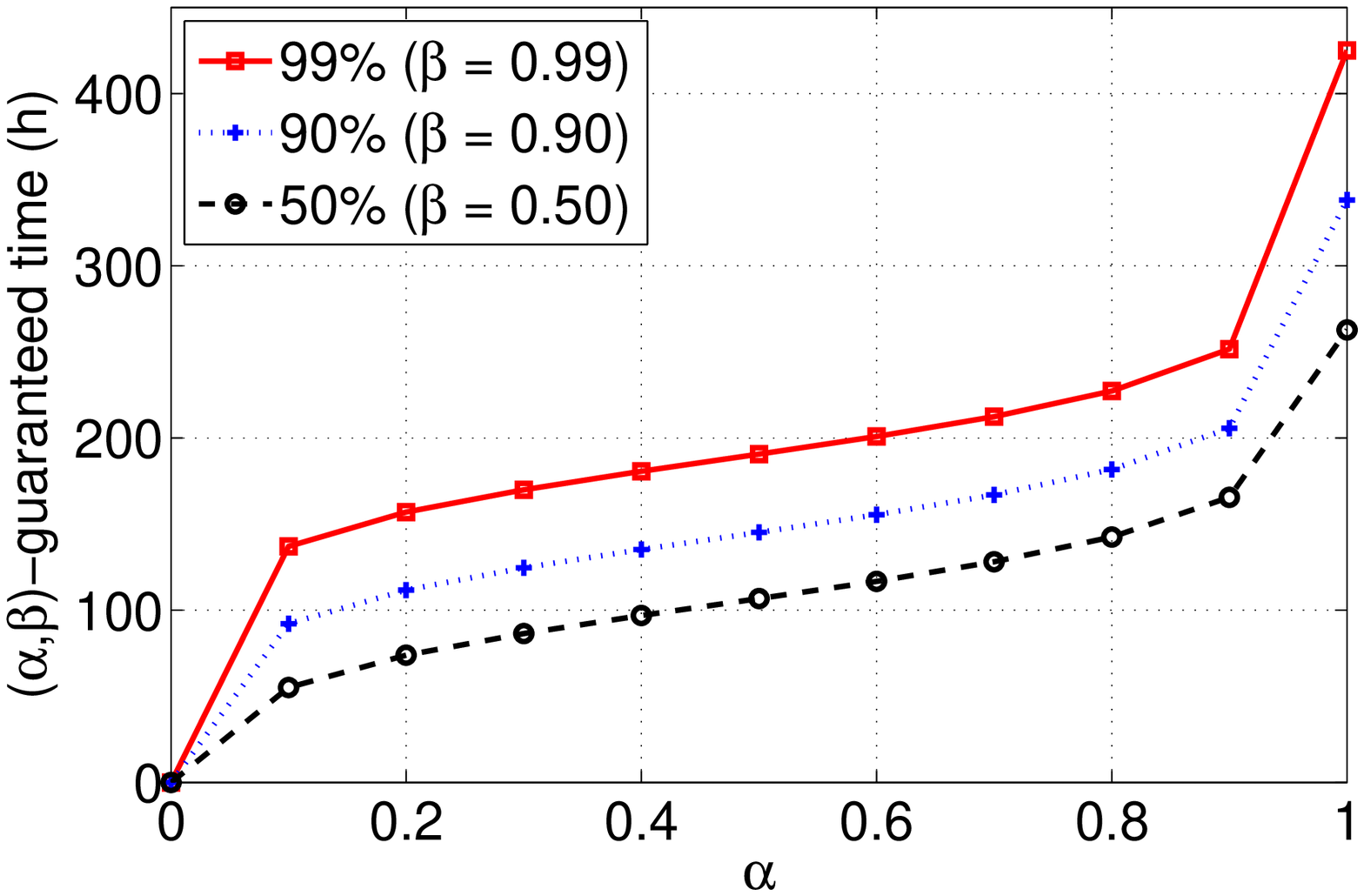,width=0.26\textwidth}}
\hspace{0.3cm}
\subfigure[$(\alpha, \beta)$-guaranteed time with 10 seeds] {\epsfig{figure=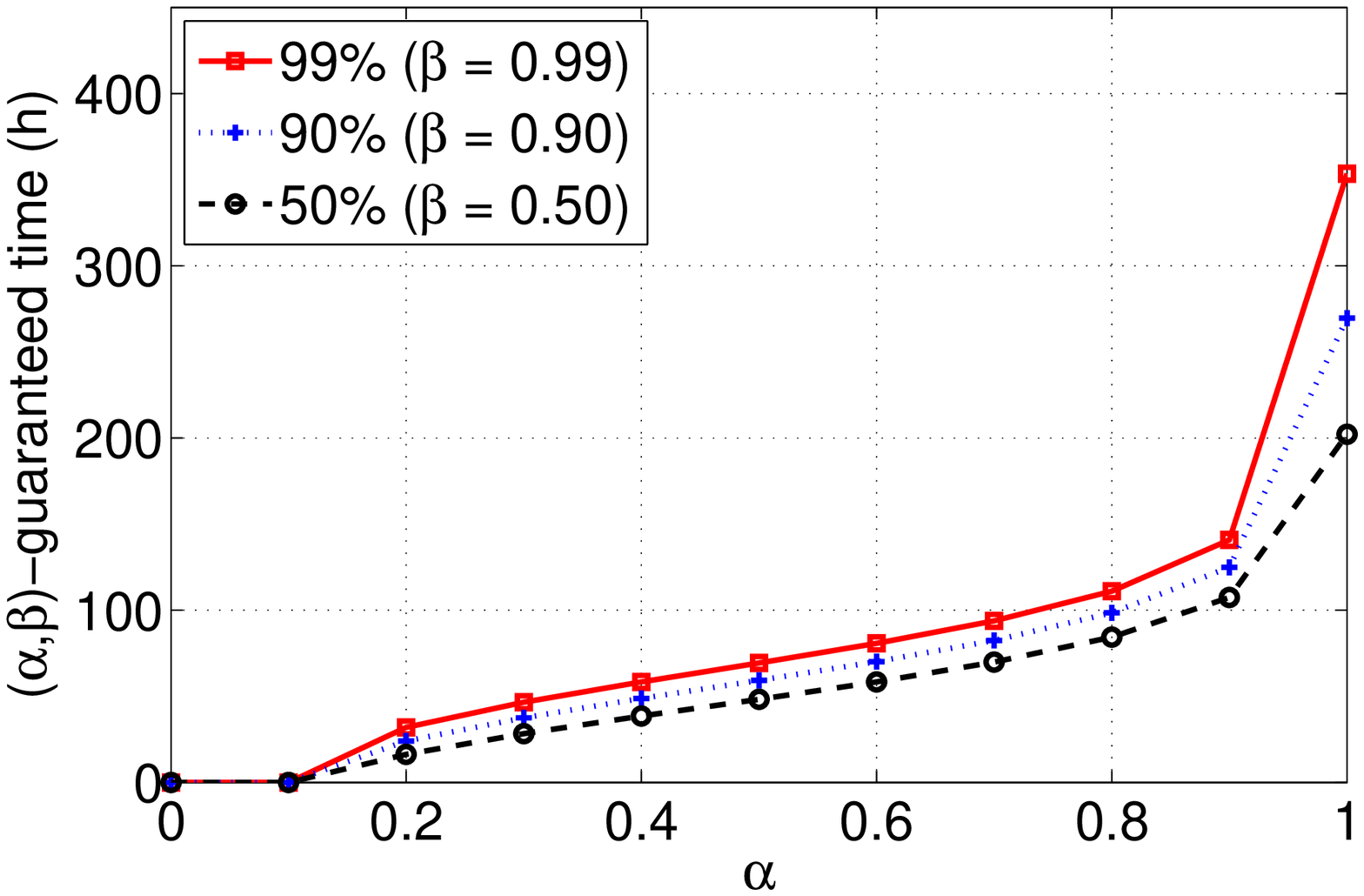, width=0.26\textwidth}}
\hspace{0.3cm}
\subfigure[$(\alpha, \beta)$-guaranteed time with 20 seeds] {\epsfig{figure=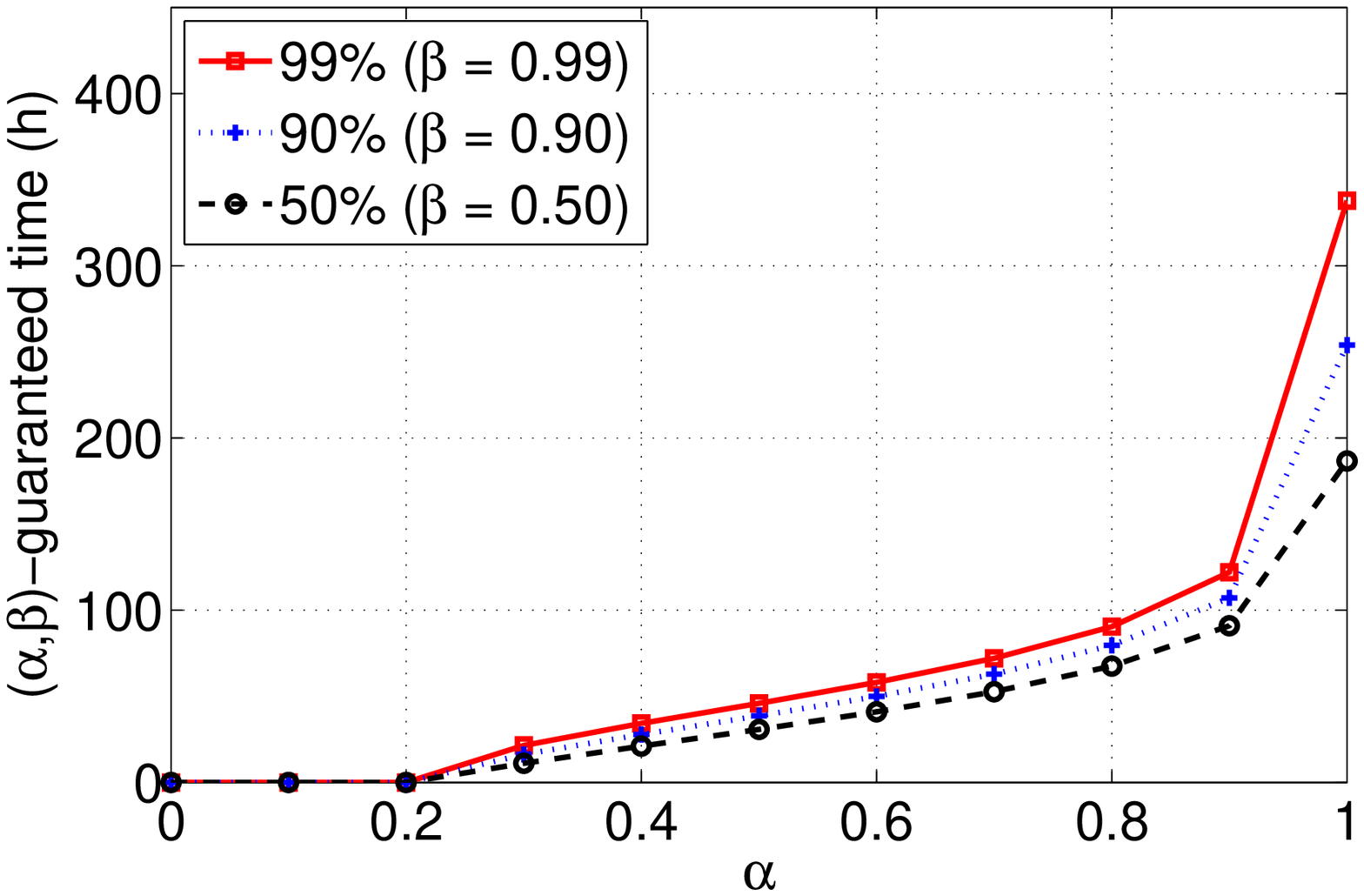, width=0.26\textwidth}}
\caption{Distribution of the $(\alpha, \beta)$-guaranteed time for $\alpha \in [0, 1]$ and $\beta = \{0.5, 0.90, 0.99\}$  with (a) 1 seed, (b) 10 seeds, and (c) 20 seeds in a homogeneous network and with (d) 1 seed, (e) 10 seeds, and (f) 20 seeds in a heterogeneous network with two groups.}
\label{fig:ctime}
\end{figure*}

We study the efficacy of our framework and characterizations using by far the largest vehicular mobility trace obtained from more than a thousand taxies in Shanghai, China~\cite{ShanghaiTaxi}.\footnote{Our framework is applicable to various networks including taxi networks. Due to the availability of data, we limit simulation study to a taxi network.} The experimental trace tracked GPS coordinates of taxies at every 30 seconds during 28 days in Shanghai. The trace was analyzed in~\cite{lee10} and it was shown that the taxies have exponentially distributed pairwise inter-contact time, which is well aligned with our CTMC-based framework.

Figs.~\ref{fig:shanghai} (a), (b), and (c) characterize the statistics of the taxi network with 1000 randomly chosen taxies in the aspect of \textit{number of contacts}, \textit{number of neighbors in a communication range} (50 meter in our analysis), and \textit{contact duration}, respectively. We apply these three factors for evaluating the effective contact rates $\lambda_{a,b}^\text{eff}= \lambda_{a,b}\varphi_{a} \psi_{b}$ derived in~\ref{eqn:exp meeting time:eff}, where $\varphi_{a} = 1$ and $\psi_b$ is 1 over average number of neighbors multiplied by the expected number of contacts to make a successful data transfer. Note that the latter is derived from the contact time distribution and the time required for a data transfer. The results for a homogeneous network (i.e., $\lambda^\star$) and for a heterogeneous network with two groups (i.e., $\lambda_{1,1}^\star$, $\lambda_{2,2}^\star$ and $\lambda_{1,2}^\star$) are summarized in Table~\ref{tbl:lambda}. Note that the infection rates in Table~\ref{tbl:lambda} satisfy the constraint in~(\ref{eqn:constraint}) that was introduced for a fair comparison between a homogeneous model and a heterogeneous model. Based on the statistics in Table~\ref{tbl:lambda}, we can predict the information spread time and examine possible methods to properly allocate resources for the taxi network.

\begin{table}[h]
\caption{Infection rates for a homogeneous network and for a heterogeneous network with two groups of taxies.}
\centering
  \begin{tabular}{|c|c|c|c|} \hline
    Homogeneous Network & \multicolumn{3}{c|}{Heterogeneous Network} \\ \hline \hline
     $ \lambda^\star$  &  $\lambda_{1,1}^\star$   & $\lambda_{2,2}^\star$   & $\lambda_{1,2}^\star \,(=\lambda_{2,1}^\star)$    \\ \hline
  $4.14 \cdot 10^{-4}$ & $7.17 \cdot 10^{-4}$ & $1.93 \cdot 10^{-4}$ & $3.72 \cdot 10^{-4}$ \\\hline
  \end{tabular}
\label{tbl:lambda}
\end{table}

Based on Table~\ref{tbl:lambda}, we simulate probabilistic guarantees for the completion time in a homogeneous and a heterogeneous network, each with 100 taxies. We assume a firmware update to be distributed for mobile devices, which will take around 90 seconds demanding 1.15 number of contacts on average. The number of taxi is scaled down to 100 due to computation complexity involved in matrix operations. Figs.~\ref{fig:ctime} (a), (b), and (c) show the $(\alpha, \beta)$-guaranteed time for $\alpha\in[0,1]$ and $\beta\in\{0.5, 0.9, 0.99\}$ with the number of seeds given by 1, 10, and 20, respectively. The figures tell that if we target 90\% penetration with 99\% confidence (i.e., $(\alpha, \beta) = (0.9, 0.99)$), then the network with a single seed is estimated to take about 11.6 days (i.e., 278 hours) to achieve the target level of information spread. This estimation largely differs from the existing estimation of average time to achieve 90\% of penetration, which is close to 7 days. This clarifies that designing plans associated with the successful spread to 90\% of nodes should allow about 4.6 days more. If not, a set of planed work may not be executable on time. If shorter time duration needs to be guaranteed to avoid the plan being delayed, our framework is able to suggest to add seeds to the network as shown in Figs.~\ref{fig:ctime} (b) and (c). As the number of seeds increases to 10 or 20, the time for 90\% penetration with 99\% confidence reduces from 278 hours to 137 and 113 hours, respectively. These predictions guide how to optimally plan the information spread.

Similarly, we can study a heterogeneous network with two groups. Figs.~\ref{fig:ctime} (d), (e), and (f) show the $(\alpha, \beta)$-guaranteed time for $\alpha \in [0,1]$ and $\beta \in \{0.5, 0.9, 0.99\}$ with 1, 10, and 20 seeds, respectively. Direct comparison between Figs.~\ref{fig:ctime} (a), (b), (c) and Figs.~\ref{fig:ctime} (d), (e), (f) confirms our claims from Theorem~\ref{thm:impact of multiple community} that the $(\alpha, \beta)$-guaranteed time in a heterogeneous network is faster for lower $\alpha$, but is slower for higher $\alpha$ close to 1. This implies that if it is mandatory to achieve 100\% penetration, making the nodes in a network to be more homogeneous (by providing more resources to relatively inactive nodes) can be helpful, when increasing the level of average contact rates is not possible due to resource concern.

\section{Conclusion}\label{sec:conclusion}
In this paper, we characterize the probabilistic guarantee of the time for information spread in opportunistic networks by developing a CTMC-based analytical framework and introducing the metric $G_{\alpha, \beta}$. We also identify the temporal scaling behavior of information spread for a set of key spread factors. Through various examples of application scenarios and simulations over the Shanghai taxi trace, we show that our framework enables us to estimate proper amount of resource to a network in information spread by providing the detailed statistics of the guaranteed time for given penetration targets. We believe our framework can be viewed as an important first step in the design of highly sophisticated acceleration methods for information spread (or prevention methods for epidemics).

\appendices
\begin{figure*}[!b]
\normalsize
\hrulefill
\vspace*{1pt}\\
\begin{align}\label{eqn:app:explicit F}
\bm{F} = \left[
\begin{matrix}
-(N-1)\lambda^\star & (N-1)\lambda^\star   & 0                    & \ldots & 0                    & 0 \\
                  0 & -2(N-2)\lambda^\star & 2(N-2)\lambda^\star  & \ldots & 0                    & 0 \\
                  0 & 0                    & -3(N-3)\lambda^\star & \ldots & 0                    & 0 \\
             \vdots & \vdots               & \vdots               & \ddots & \vdots               & \vdots \\
                  0 & 0                    & 0                    & \ldots & -2(N-2)\lambda^\star & 2(N-2)\lambda^\star \\
                  0 & 0                    & 0                    & \ldots & 0                    & -(N-1)\lambda^\star
\end{matrix}\right].
\end{align}
\end{figure*}
\section{Proof of Equation~(\ref{eqn:exp meeting time:eff})}\label{sec:appendix:proof of eff meeting time}
For a given $t\ge0$, let $N_{a,b}(t)$ be the total number of contacts between nodes~$a$ and~$b$ by time~$t$. Since $\{N_{a,b}(t); t \ge 0\}$ is a Poisson process with rate $\lambda_{a,b}$, we have for $n=0,1,\ldots$:
\begin{align}\label{eqn:app:eff meeting:1}
\pr\{N_{a,b}(t)=n\} = \exp(-\lambda_{a,b}t)\frac{(\lambda_{a,b}t)^n}{n!}.
\end{align}
Since a contact between nodes~$a$ and $b$ incurs infection with probability $\varphi_a \psi_b$, we have for $n=0,1,\ldots$ the following:
\begin{align}\label{eqn:app:eff meeting:2}
\pr\{M_{a,b}^\text{eff} > t\,|\, N_{a,b}(t) = n\} = (1-\varphi_a \psi_b)^n.
\end{align}
Hence, from~(\ref{eqn:app:eff meeting:1}) and~(\ref{eqn:app:eff meeting:2}), $\pr\{M_{a,b}^\text{eff} > t\}$ is obtained by:
\begin{align*}
\pr\{\!M_{a,b}^\text{eff} > t\}
&= \sum_{n=0}^{\infty}\pr\{\!M_{a,b}^\text{eff} > t\,|\, N_{a,b}(t) = n\} \pr\{\!N_{a,b}(t)\! = \!n\}\! \\
&= \exp(-\lambda_{a,b}t) \sum_{n=0}^{\infty}\frac{\big((1-\varphi_a \psi_b)\lambda_{a,b}t\big)^n}{n!} \\
&= \exp(-\lambda_{a,b}t) \exp((1-\varphi_a \psi_b)\lambda_{a,b}t) \\
&= \exp(-\varphi_a \psi_b\lambda_{a,b}t).
\end{align*}

\section{Proof of Lemma~\ref{lemma:CTMC model}}\label{sec:appendix:proof of lemma 1}
The derivations in Examples~I and II are easily extended to prove the main result, (P1), and (P2). The proof of (P3) is given in~\cite{bremaud08}. Hence, we omit the details and provide intuition of (P4). Suppose $\bm{I}(t) = \bm{e}_i$. Since $\{\sum_{k}I_k(t); t \ge 0\}$ is a counting process, state transitions occur to an adjacent state~$\bm{e}_j$ stratifying $\sum_{k}|j_k-i_k|=1$ and $j_l=i_l+1$ for some~$l$. In addition, the time required to transit to such state $\bm{e}_j$ becomes
\begin{align}\label{eqn:app:lemma1:rate}
\min\big\{M_{a,b}^{\text{eff}}\, ; a\in  \mathcal{I}_1(t) \cup \ldots \cup \mathcal{I}_K(t),  b \in \mathcal{S}_l(t)\big\},
\end{align}
where $\mathcal{I}_k(t)$ and $\mathcal{S}_k(t)\,(k=1,\ldots,K)$ denote index sets of infected nodes and susceptible nodes in group~$k$ at time~$t$, respectively. By~(\ref{eqn:exp meeting time:eff}) and the independence of $M_{a,b}^{\text{eff}}$, the random variable in~(\ref{eqn:app:lemma1:rate}) follows an exponential distribution with rate $(N_l-i_l)\sum_{k}i_k \lambda^\star_{k,l}$. This yields (P4).

\section{Explicit Expression for Fundamental Matrix~$\bm{F}$}\label{sec:appendix:fundamental matrix}
In this appendix, we show explicitly the fundamental matrices $\bm{F}$ for $K=1,2$ in Examples I and~II. Suppose $K=1$. Then, from Fig.~\ref{fig:transition diagram when K=1}, we can obtain the matrix~$\bm{F}$ as given at the bottom of the page. Suppose $K=2$. Then, from Fig.~\ref{fig:transition diagram when K=2}, the matrix~$\bm{F}$ is obtained as follows: for $i=0,1,\ldots,N_1$, define matrices $\bm{A}_i$ and $\bm{B}_i$ as
\begin{align*}
\bm{A}_{i} \deq \left[
\begin{matrix}
-z_{i,0} &  x_{i,0} & 0        & \ldots & 0                & 0 \\
0        & -z_{i,1} &  x_{i,1} & \ldots & 0                & 0 \\
0        & 0        & -z_{i,2} & \ldots & 0                & 0 \\
\vdots   & \vdots   & \vdots   & \ddots & \vdots           & \vdots \\
0        & 0        & 0        & \ldots & -z_{i,N_2-1}     &  x_{i,N_2-1} \\
0        & 0        & 0        & \ldots & 0                & -z_{i,N_2} \\
\end{matrix}\right],
\end{align*}
\begin{align*}
\bm{B}_{i} \deq \left[
\begin{matrix}
y_{i,0}  & 0        & 0        & \ldots & 0                & 0 \\
0        & y_{i,1}  & 0        & \ldots & 0                & 0 \\
0        & 0        & y_{i,2}  & \ldots & 0                & 0 \\
\vdots   & \vdots   & \vdots   & \ddots & \vdots           & \vdots \\
0        & 0        & 0        & \ldots & y_{i,N_2-1}      & 0 \\
0        & 0        & 0        & \ldots & 0                & y_{i,N_2} \\
\end{matrix}\right],
\end{align*}
where the components $x_{i,j}, y_{i,j}$, and $z_{i,j}$ are defined as
\begin{align*}
x_{i,j} &\deq i(N_2-j)\lambda^\star_{1,2} + j(N_2-j)\lambda^\star_{2,2},\\
y_{i,j} &\deq i(N_1-i)\lambda^\star_{1,1} + j(N_1-i)\lambda^\star_{2,1},\\
z_{i,j} &\deq x_{i,j} + y_{i,j}.
\end{align*}
Then, the fundamental matrix~$\bm{F}$ is given by
\begin{align*}
\bm{F} = \left[
\begin{matrix}
\bm{\tilde{A}}_{0}   & \bm{\tilde{B}}_{0}  & \bm{0}       & \ldots & \bm{0} & \bm{0} \\
\bm{0}       & \bm{A}_{1}   & \bm{B}_{1}   & \ldots & \bm{0} & \bm{0} \\
\bm{0}       & \bm{0}       & \bm{A}_{2}   & \ldots & \bm{0} & \bm{0} \\
\vdots       & \vdots       & \vdots       & \ddots & \vdots & \vdots \\
\bm{0}       & \bm{0}       & \bm{0}       & \ldots & \bm{A}_{N_1-1} & \bm{\hat{B}}_{N_1-1} \\
\bm{0}       & \bm{0}       & \bm{0}       & \ldots & \bm{0} & \bm{\hat{A}}_{N_1} \\
\end{matrix}\right],
\end{align*}
where $\bm{\tilde{A}}_0$ is obtained by eliminating the first row and the first column of $\bm{A}_0$, and $\bm{\tilde{B}}_0$ is obtained by eliminating the first row of $\bm{B}_0$ (here, the elimination is for excluding state transition from $(0,0)$ or to $(0,0)$). $\bm{\hat{A}}_{N_1}$ is obtained by eliminating the last row and the last column of $\bm{A}_{N_1}$, and $\bm{\hat{B}}_{N_1-1}$ is obtained by eliminating the last column of $\bm{B}_{N_1-1}$ (here, the elimination is for excluding state transition from $(N_1,N_2)$ or to $(N_1,N_2)$).

\section{Proof of Lemma~\ref{lemma:distribution of completion time}}\label{sec:appendix:proof of lemma 2}
Since the event $\{T_\alpha > t\}$ is equivalent to $\{\bm{I}_\alpha(t)\in\mathcal{E}_\alpha^\star\}$, we have
\begin{align}\label{eqn:app:lemma2:ccdf1}
\pr\{T_\alpha > t\} = \pr\{\bm{I}_\alpha(t)\in\mathcal{E}_\alpha^\star\}.
\end{align}
Let $\bm{\pi}_\alpha(t) \deq (\pr\{\bm{I}_\alpha(t)=\bm{e}\})_{\bm{e}\in\mathcal{E}_\alpha^\star}$ be the distribution of $\bm{I}_\alpha(t)$ on $\mathcal{E}_\alpha^\star$. Then, by the same reason in (P3) of Lemma~\ref{lemma:CTMC model}, we have
\begin{align*}
\bm{\pi}_\alpha(t) = \bm{\pi}_\alpha(0)\exp(\bm{F}_\alpha t) = \bm{h}_\alpha \exp(\bm{F}_\alpha t).
\end{align*}
Hence, the probability $\pr\{\bm{I}_\alpha(t)\in\mathcal{E}_\alpha^\star\}$ is derived as
\begin{align}\label{eqn:app:lemma2:ccdf2}
\pr\{\bm{I}_\alpha(t)\in\mathcal{E}_\alpha^\star\} = |\bm{\pi}_\alpha(t)| = \bm{h}_\alpha \exp(\bm{F}_\alpha t) \bm{1}.
\end{align}
By combining (\ref{eqn:app:lemma2:ccdf1}) and (\ref{eqn:app:lemma2:ccdf2}), the CDF $H_\alpha(t) (\deq \pr\{T_\alpha \le t\})$ is obtained as
\begin{align*}
H_\alpha(t) = 1-\pr\{T_\alpha > t\}= 1-\bm{h}_\alpha \exp(\bm{F}_\alpha t) \bm{1},
\end{align*}
which proves the first formula in Lemma~\ref{lemma:distribution of completion time}.

From basic Markov chain theory, the CDF $H_\alpha(t)$ can be expressed as follows~\cite[Eq.~(1)]{aalen95}:
\begin{align}\label{eqn:app:lemma2:ccdf3}
H_\alpha(t) = 1-\sum_{i}\exp(\rho_i t) P_{i}(t),
\end{align}
where $\rho_i$ denote nonzero eigenvalues of~$\bm{Q}_{\alpha}$. Since $\{\sum_{k}I_{k}(t); t\ge 0\}$ is a counting process, the infinitesimal generator $\bm{Q}_{\alpha}$ is an upper triangular matrix. Hence, all the nonzero eigenvalues of~$\bm{Q}_{\alpha}$ come from the diagonal elements of the matrix~$\bm{F}_\alpha$, which are real and negative. Hence, (\ref{eqn:app:lemma2:ccdf3}) can be rewritten as
\begin{align*}
H_\alpha(t) = 1-\sum_{i}\exp(-|\rho_i| t) P_{i}(t),
\end{align*}
which proves the second formula in Lemma~\ref{lemma:distribution of completion time}.

\section{Proof of Lemma~\ref{lemma:formula for metrics}}\label{sec:appendix:proof of lemma 3}
It is clear from the formulas in Lemma~\ref{lemma:distribution of completion time} that the function~$H_\alpha(\cdot)$ is strictly increasing. Hence, the $(\alpha,\beta)$-guaranteed time~$G_{\alpha,\beta}$ is uniquely determined by solving~$H_{\alpha}(G_{\alpha,\beta})=\beta$. Since $H_\alpha(\cdot)$ is a bijective function, it has the inverse function $H_{\alpha}^{-1}(\cdot)$. Therefore,
$G_{\alpha,\beta}$ is obtained by $G_{\alpha,\beta} = H_\alpha^{-1}(\beta)$. This proves~(\ref{eqn:lemma:guaranteed time}).

In our model, the Markov chain $\{\bm{I}_\alpha(t); t \ge 0\}$ is eventually absorbed into the absorbing state space $\mathcal{E}_\alpha^o$ with probability~1, which shows the existence of the inverse matrix of~$\bm{F}_\alpha$~\cite[Lemma~2.2.1.]{neuts81},~\cite[Theorem~2.4.3]{Latouche99}. Under this condition, it is well-known that the $n$th moment of $T_\alpha$ is given by~(\ref{eqn:lemma:nth moment})~\cite[Eq. (2.2.7)]{neuts81},\cite[Eq. (2.13)]{Latouche99}, which completes the proof.

\section{Proof of Theorem~\ref{thm:impact of lambda level}}\label{sec:appendix:proof of thm 1}
Suppose that the infection rate $\lambda_{a,b}^{\text{eff}}$ is scaled by $\gamma\,(>0)$ times for all $a,b$. In this proof, we add a symbol $\hat{}$ on top of any notation to distinguish it after the scale. Note that by (P4) of Lemma~\ref{lemma:CTMC model} and~(\ref{eqn:truncated F}), we have $\bm{\hat{F}}_\alpha = \gamma\bm{F}_\alpha$. Hence, by Lemma~\ref{lemma:distribution of completion time} we have for all $t\ge 0$:
\begin{align}\label{eqn:thm:lambda level 1}
\pr\{\hat{T}_\alpha \le t\}
&= 1-\bm{h}_\alpha \exp(\bm{\hat{F}}_\alpha t) \bm{1} \nonumber \\
&= 1-\bm{h}_\alpha \exp(\bm{F}_\alpha (\gamma t)) \bm{1} \nonumber \\
&= \pr\{T_\alpha \le \gamma t\}.
\end{align}
That is, $\pr\{\hat{T}_\alpha \le t\} = \pr\{\gamma^{-1} T_\alpha \le t\}$ for all $t\ge 0$. This proves~(\ref{eqn:thm:lambda level}).

Let $\hat{H}_\alpha(t) \deq \pr\{\hat{T}_\alpha \le t\}$ be the CDF of $\hat{T}_\alpha$. Then,~(\ref{eqn:thm:lambda level 1}) gives $\hat{H}_\alpha(t) = H_\alpha(\gamma t)$, which yields $\hat{H}^{-1}_\alpha(t) = \gamma^{-1}H_\alpha^{-1}( t)$. By~(\ref{eqn:lemma:guaranteed time}) in Lemma~\ref{lemma:formula for metrics}, we further have
\begin{align*}
\hat{G}_{\alpha,\beta} = \hat{H}^{-1}_\alpha(\beta) = \gamma^{-1}H_\alpha^{-1}(\beta) = \gamma^{-1}G_{\alpha,\beta}.
\end{align*}
This proves~$\hat{G}_{\alpha,\beta}= \gamma^{-1}G_{\alpha,\beta}$. From~(\ref{eqn:thm:lambda level}), we have $(\hat{T}_\alpha)^n \ed \gamma^{-n}(T_\alpha)^n$. By taking expectation, $E[(\hat{T}_\alpha)^n] = \gamma^{-n} E[(T_\alpha)^n]$. This completes the proof.

\section{Proof of $\hat{\mathcal{M}}(t)=\mathcal{M}(\gamma t)$ and $\hat{\mathcal{D}}(t)=\gamma\mathcal{D}(\gamma t)$}\label{sec:appendix:proof of avg scaling}
Similarly to the proof of Theorem~\ref{thm:impact of lambda level}, we add a symbol $\hat{}$ on top of any notation to distinguish it after the scale. Since the random variable $\sum_{k}I_k(t)$ takes on only nonnegative integer values from 0 to $N$, the expectation $\mathcal{M}(t)\,(\deq \E[\sum_{k} I_k(t)])$ can be obtained by $\mathcal{M}(t) = \sum_{i=1}^{N} \pr\{\sum_{k} I_k(t) \ge i\}.$ By Definition~\ref{def:alpha completion time}, the event $\{\sum_{k} I_k(t) \ge i\}$ is equivalent to $\{T_{i/N} \le t\}$. Hence, the expectation $\mathcal{M}(t)$ is given by
\begin{align}\label{eqn:app:avg scaling:1}
\mathcal{M}(t) = \sum_{i=1}^{N} \pr\{T_{i/N} \le t \}.
\end{align}
Similarly, the expectation $\hat{\mathcal{M}}(t)$ after the scale is given by
\begin{align}\label{eqn:app:avg scaling:2}
\hat{\mathcal{M}}(t) = \sum_{i=1}^{N} \pr\{\hat{T}_{i/N} \le t \}.
\end{align}
By (\ref{eqn:thm:lambda level}) in Theorem~\ref{thm:impact of lambda level}, the probability in~(\ref{eqn:app:avg scaling:2}) satisfies $\pr\{\hat{T}_{i/N} \le t \} = \pr\{T_{i/N} \le \gamma t \}$. Thus, from (\ref{eqn:app:avg scaling:1}) we have
\begin{align*}
\hat{\mathcal{M}}(t) = \sum_{i=1}^{N} \pr\{T_{i/N} \le \gamma t \} = \mathcal{M}(\gamma t).
\end{align*}
By using the relation $\hat{\mathcal{M}}(t)=\mathcal{M}(\gamma t)$, it is straightforward to show that $\hat{\mathcal{D}}(t)\,(\deq \frac{\text{d}}{\text{d} t} \hat{\mathcal{M}}(t))=\gamma\mathcal{D}(\gamma t)$.

\section{Proof of Theorem~\ref{thm:impact of population size}}\label{sec:appendix:proof of thm 2}
In this appendix, we will prove the followings in order:
\begin{enumerate}[(T1)]
\item $G_{\alpha,\beta}$ is strictly decreasing with $N$ for sufficiently large~$\beta$.
\item $\E[T_\alpha]$ is strictly decreasing with $N$.
\item $G_{\alpha,\beta} = \Theta\big((\lambda^\star)^{-1} N^{-1}(\log N-\log(\log\frac{1}{\beta}))\big).$
\item $\E[T_\alpha] = \Theta\big((\lambda^\star)^{-1} N^{-1}\log N\big).$
\end{enumerate}
In the proof, we add the subscript $N$ to the variables $T_\alpha$ and $G_{\alpha,\beta}$ to explicitly denote the assumed population size.

\begin{figure}[t!]
\centering
{\epsfig{figure=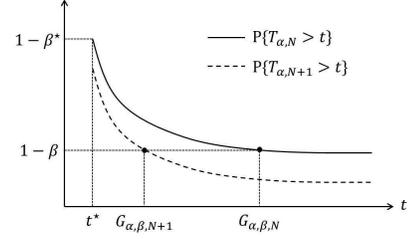,width=0.3\textwidth}}
\caption{Proof of (T1) in Theorem~\ref{thm:impact of population size}: $G_{\alpha,\beta,N} > G_{\alpha,\beta,N+1}$ for any $\beta \ge \beta^\star$.}
\label{fig:illustration}
\end{figure}

\smallskip\smallskip
\noindent\textit{Proof of (T1):} When $\alpha=1$, we have $\bm{F}_\alpha = \bm{F}$, and accordingly all the eigenvalues of $\bm{F}_\alpha$ come from the diagonal elements of the matrix~$\bm{F}$. In addition, when $K=1$, the diagonal elements of~$\bm{F}$ can be obtained from~(\ref{eqn:app:explicit F}) by $\rho_i = -i(N-i)\lambda^\star, i=1,2,\ldots,\lfloor \frac{N}{2} \rceil$. Hence, by Lemma~\ref{lemma:distribution of completion time}, we have
\begin{align}\label{eqn:app:thm2:ccdf 1}
\pr\{T_{\alpha,N} > t\} = \sum_{i=1}^{\lfloor \frac{N}{2}\rfloor} \exp(-i(N-i)\lambda^\star t)P_{i,N}(t).
\end{align}
Similarly, for a network with $N+1$ nodes, we have
\begin{align}\label{eqn:app:thm2:ccdf 2}
\pr\{T_{\alpha,N+1} \!>\! t\} \!=\!\!\! \sum_{i=1}^{\lfloor \frac{N+1}{2}\rfloor} \!\! \exp(-i(N\!+\!1\!-\!i)\lambda^\star t)P_{i,N+1}(t).
\end{align}
It is straightforward to show that the ratio of (\ref{eqn:app:thm2:ccdf 2}) to (\ref{eqn:app:thm2:ccdf 1}) converges as
\begin{align*}
\lim_{t\to\infty}\frac{\pr\{T_{\alpha,N+1} > t\}}{\pr\{T_{\alpha,N} > t\}}=0.
\end{align*}
Thus, there exists $t^\star\,(\ge 0)$ such that $\pr\{T_{\alpha,N+1}>t\} < \pr\{T_{\alpha,N}>t\}$ for all $t \ge t^\star$. Let $\beta^\star \deq H_\alpha(t^\star)$. Then, as evident by Fig.~\ref{fig:illustration}, we have $G_{\alpha,\beta,N} > G_{\alpha,\beta,N+1}$ for any $\beta \ge \beta^\star$.

\smallskip\smallskip
\noindent\textit{Proof of (T2):} From Fig.~\ref{fig:transition diagram when K=1}, the expectation $\E[T_{\alpha,N}]$ is obtained by $\E[T_{\alpha,N}] = \frac{1}{\lambda^\star} \sum_{i=1}^{N-1} \frac{1}{i(N-i)}$. Similarly, for a network with $N+1$ nodes, we have $\E[T_{\alpha,N+1}] = \frac{1}{\lambda^\star} \sum_{i=1}^{N} \frac{1}{i(N+1-i)}$. By using these formulas for $\E[T_{\alpha,N}]$ and $\E[T_{\alpha,N+1}]$, we can show that $\E[T_{\alpha,N+1}]-\E[T_{\alpha,N}]<0$ for all $N=1,2,\ldots$ as follows:
\begin{align}\label{eqn:app:thm2:increasing mean}
&\E[T_{\alpha,N+1}]-\E[T_{\alpha,N}] \nonumber \\
&\quad = \frac{1}{\lambda^\star} \bigg(\sum_{i=1}^{N}\frac{1}{i(N+1-i)} - \sum_{i=1}^{N-1} \frac{1}{i(N-i)}\bigg) \nonumber \\
&\quad = \frac{1}{\lambda^\star} \bigg(\frac{1}{N}+\sum_{i=1}^{N-1}\frac{1}{N-i}\Big(\frac{1}{i+1} - \frac{1}{i}\Big)\bigg) \nonumber \\
&\quad < \frac{1}{\lambda^\star} \bigg(\frac{1}{N}+\frac{1}{N-1}\sum_{i=1}^{N-1}\Big(\frac{1}{i+1} - \frac{1}{i}\Big)\bigg) \nonumber \\
&\quad = \frac{1}{\lambda^\star} \bigg(\frac{1}{N}-\frac{1}{N}\bigg) =0,
\end{align}
which proves (T2).

\smallskip\smallskip
\noindent\textit{Proof of (T3):} To prove (T3), we need the following lemma.
\begin{lemma}\label{lemma:hypoexponential}
Let $Z_i\,(i=1,\ldots,n)$ be a sequence of independent exponential random variables with rates $\eta i$. Then, the sum $Z\deq \sum_{i=1}^{n}Z_i$ has the complementary cumulative distribution function~(CCDF) give by:
\begin{align*}
\pr\{Z > z\} = \sum_{i=1}^{n} (-1)^{i-1} {n \choose i}\exp(- \eta i z).
\end{align*}
\end{lemma}

\noindent\textit{Proof:} It is well-known that the sum of $n$ independent exponential random variables with rates $r_i\,(i=1,\ldots,n)$ follows the generalized Erlang distribution. When $r_i \neq r_j$ for all $i \neq j$, i.e., in our case $r_i = \eta i $, the CCDF of the generalized Erlang distribution is given by
\begin{align}\label{eqn:app:lemma4:hypo}
\pr\{Z > z\} = \sum_{i=1}^{n} \bigg(\prod_{j=1,j\neq i}^{n}\frac{r_j}{r_j-r_i}\bigg) \exp(- r_i z).
\end{align}
Replacing $r_i$ by $\eta i$ and simplifying (\ref{eqn:app:lemma4:hypo}) yield the lemma. \hfill $\blacksquare$

\smallskip\smallskip
Suppose that $N$ is an odd number. When $N$ is an even number, we can use similar steps for the proof, and hence we only consider the case when $N$ is odd. From Fig.~\ref{fig:transition diagram when K=1}, the $\alpha$-completion time is obtained by
\begin{align}\label{eqn:app:thm2:T_alpha}
T_{\alpha,N} = \sum_{i=1}^{\frac{N-1}{2}} (X_i + Y_i),
\end{align}
where $X_i$ and $Y_i$ are independent and identically distributed exponential random variables with rates~$i(N-i)\lambda^\star$. To give a bound on $T_{\alpha,N}$, we introduce random variables $T_{\text{upper}}\deq \sum_{i=1}^{N-1}Z_i^u$ and $T_{\text{lower}}\deq \sum_{i=1}^{N-1}Z_i^l$, where $Z_i^u$ and $Z_i^l$ are independent exponential random variables with rates $\lambda^\star N i/4$ and $\lambda^\star N i$, respectively. For any two random variables $A$ and $B$, we use $A \preceq B$ to denote $\pr\{A > x\} \le \pr\{B > x\}$ for all $x \in\mathbb{R}$. In the following, we show that
\begin{align}\label{eqn:app:thm2:st order}
T_{\text{lower}} \preceq T_\alpha \preceq T_{\text{upper}}.
\end{align}
Since the rate of $X_i$ is greater than that of $Z_{2i}^u$, we have $X_i \preceq Z_{2i}^u$ for $i=1,2,\ldots,\frac{N-1}{2}$. In addition, since $Z_{2i}^u \preceq Z_{2i-1}^u$ by the same reason and $X_i \ed Y_i$, we have $Y_i \preceq Z_{2i-1}^u$ for $i=1,2,\ldots,\frac{N-1}{2}$. Therefore, we have $\sum_{i=1}^{\frac{N-1}{2}} (X_i + Y_i) \preceq  \sum_{i=1}^{N-1}Z_i^u$. That is, $T_\alpha \preceq T_{\text{upper}}$. By using a similar approach as before, we can easily obtain $T_{\text{lower}}\preceq T_\alpha$. Due to similarity, we omit the details.

Let $t_{\text{upper}}^\star \deq \frac{4}{\lambda^\star N}\{\log(N-1)-\log(\log\frac{1}{\beta})\}$, and $t_{\text{lower}}^\star \deq \frac{1}{4}t_{\text{upper}}^\star$. Then, we have the followings, which will be shown in the sequel:
\begin{align}\label{eqn:app:thm3:order}
\begin{split}
\lim_{N\to \infty} \pr\{T_{\text{lower}} > t_{\text{lower}}^\star\} &= 1-\beta, \\
\lim_{N\to \infty} \pr\{T_{\text{upper}} > t_{\text{upper}}^\star\} &= 1-\beta.
\end{split}
\end{align}
The results in (\ref{eqn:app:thm2:st order}) and (\ref{eqn:app:thm3:order}) show that there exists $N^\star \in\mathbb{N}$ such that
\begin{align*}
t_{\text{lower}}^\star \le G_{\alpha,\beta,N} \le t_{\text{upper}}^\star \quad \text{for all } N \ge N^\star,
\end{align*}
which gives $G_{\alpha,\beta,N} = \Theta((\lambda^\star)^{-1} N^{-1}(\log N-\log(\log\frac{1}{\beta}))).$

It remains to prove (\ref{eqn:app:thm3:order}). By Lemma~\ref{lemma:hypoexponential}, the CCDF of $T_{\text{lower}}$ is given by
\begin{align*}
\pr\{T_{\text{lower}} > t\} = \sum_{i=1}^{N-1} (-1)^{i-1} {N-1 \choose i}\exp(- \lambda^\star N i t).
\end{align*}
Hence, $\pr\{T_{\text{lower}} > t_{\text{lower}}^\star\}$ is simplified as
\begin{align}\label{eqn:app:thm3:scaling}
\pr\{T_{\text{lower}} > t_{\text{lower}}^\star\} \nonumber
&= \sum_{i=1}^{N-1} (-1)^{i-1} {N-1 \choose i}\bigg(\frac{1}{N-1}\log\frac{1}{\beta}\bigg)^i \nonumber \\
&= 1-\sum_{i=0}^{N-1} {N-1 \choose i}\bigg(\frac{\log \beta}{N-1}\bigg)^i \nonumber \\
&= 1-\bigg(1+\frac{\log \beta }{N-1}\bigg)^{N-1}.
\end{align}
By letting $N$ go to $\infty$, we have
\begin{align*}
\lim_{N\to \infty} \pr\{T_{\text{lower}} > t_{\text{lower}}^\star\} = 1-\exp(\log \beta)= 1-\beta, \end{align*}
which proves the first equality in~(\ref{eqn:app:thm3:order}). Similarly as above, we can prove the second equality in~(\ref{eqn:app:thm3:order}) and omit detailed derivations.

\smallskip\smallskip
\noindent\textit{Proof of (T4):} Suppose that $N$ is an odd number. When $N$ is an even number, we can use similar steps for the proof, and hence we only consider the case when $N$ is odd. From~(\ref{eqn:app:thm2:T_alpha}), the expectation of $T_{\alpha,N}$ is given by
\begin{align}\label{eqn:app:thm2:avg T_alpha}
\E[T_{\alpha,N}] = \frac{2}{\lambda^\star}\sum_{i=1}^{\frac{N-1}{2}}\frac{1}{i(N-i)}.
\end{align}
For notational simplicity, we define a function $f(x) \deq \frac{1}{x(N-x)}$ for $0 <x<N$. Since $f(\cdot)$ is a strictly decreasing convex function, the finite series in (\ref{eqn:app:thm2:avg T_alpha}) is bounded above as follows:
\begin{align*}
\E[T_{\alpha,N}] = \frac{2}{\lambda^\star}\sum_{i=1}^{\frac{N-1}{2}}f(i) \le \frac{2}{\lambda^\star} \bigg\{ f(1) + \int_{1}^{\frac{N-1}{2}}f(x) \,\text{d}x \bigg\}.
\end{align*}
Using basic calculus, we obtain $f(1) + \int_{1}^{\frac{N-1}{2}}f(x) \,\text{d}x = \frac{1}{N}\log\frac{(N-1)^2}{N+1}$, which gives $\E[T_{\alpha,N}] = O((\lambda^\star)^{-1} N^{-1}\log N).$ By the same reason as above, the finite series in (\ref{eqn:app:thm2:avg T_alpha}) is bounded below as follows:
\begin{align*}
\E[T_{\alpha,N}] \ge \frac{2}{\lambda^\star} \int_{1}^{\frac{N+1}{2}}f(x) \,\text{d}x = \frac{2}{\lambda^\star N}\log(N+1),
\end{align*}
which gives $\E[T_{\alpha,N}] = \Omega((\lambda^\star)^{-1} N^{-1}\log N).$ This completes the proof.

\section{Proof of Remark~\ref{rmk}}\label{sec:appendix:proof of rmk}
\begin{figure}[t!]
\centering
{\epsfig{figure=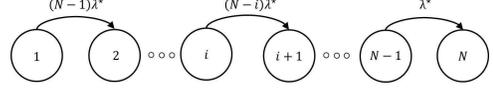,width=0.35\textwidth}}
\caption{Transition diagram of the Markov chain $\{I(t); t\ge 0\}$ when $K=1$ in the non-cooperative model.}
\label{fig:transition diagram in non-cooperative model}
\end{figure}

In the non-cooperative model, the number of infected nodes evolves as a CTMC with transition diagram depicted in Fig.~\ref{fig:transition diagram in non-cooperative model}. Let $T_{\alpha,N}^{o}$ denote the $\alpha$-completion time in a network with $N$ non-cooperative nodes. As in the proof of Theorem~\ref{thm:impact of population size}, we use the subscript $N$ to explicitly denote the assumed population size. From Fig.~\ref{fig:transition diagram in non-cooperative model}, we have
\begin{align}\label{eqn:app:rmk:T_alpha}
T_{\alpha,N}^{o} = \sum_{i=1}^{N-1}Z_i^o,
\end{align}
where $Z_i^o$ are independent exponential random variables with rates $\lambda^\star i$. From (\ref{eqn:app:rmk:T_alpha}), it is clear that $T_{\alpha,N+1}^{o} \ed T_{\alpha,N}^{o} + Z_{N}^o$, which gives
\begin{align*}
\pr\{T_{\alpha,N}^{o} > t\} <\pr\{T_{\alpha,N+1}^{o} > t\} \quad \text{for all } t>0.
\end{align*}
Therefore, the guaranteed time $G_{\alpha,\beta,N}$ is strictly increasing with~$N$. From (\ref{eqn:app:rmk:T_alpha}), it is also clear that $\E[T_{\alpha,N}^{o}] = \frac{1}{\lambda^\star}\sum_{i=1}^{N-1}\frac{1}{i}$, which is the $(N-1)$th partial sum of the harmonic series divided by $\lambda^\star$. Hence, the average $\E[T_{\alpha,N}^{o}]$ is strictly increasing with~$N$, and it scales as $\Theta((\lambda^\star )^{-1}\log N)$.

By Lemma~\ref{lemma:hypoexponential}, the CCDF of $T_{\alpha,N}^o$ is given by
\begin{align*}
\pr\{T_{\alpha,N}^o > t\} = \sum_{i=1}^{N-1} (-1)^{i-1} {N-1 \choose i}\exp\big(- \lambda^\star i t\big).
\end{align*}
Let $t^\star \deq \frac{1}{\lambda^\star}\{\log(N-1)-\log(\log\frac{1}{\beta})\}$. By using the same approach as in (\ref{eqn:app:thm3:scaling}), we can derive
\begin{align*}
\pr\{T_{\alpha,N}^o > t^\star\}
&= 1-\bigg(1+\frac{\log \beta }{N-1}\bigg)^{N-1}.
\end{align*}
Due to similarity, we omit the details. By letting $N$ go to $\infty$, we have $\lim_{N\to \infty} \pr\{T_{\alpha,N}^o > t^\star\} = 1-\exp(\log \beta)= 1-\beta,$ which proves $G_{\alpha,\beta} = \Theta((\lambda^\star )^{-1}(\log N-\log(\log\frac{1}{\beta})))$.

In the following, we will prove the results in the table. We first consider the cooperative model. Since $T_{\alpha,N}$ is obtained by the sum of independent exponential random variables with rates $i(N-1)\lambda^\star\,(i=1,2,\ldots,N-1)$ (see Fig.~\ref{fig:transition diagram when K=1}), the variance of $T_{\alpha,N}$, denoted by $\text{Var}(T_{\alpha,N})$, is derived as
\begin{align}\label{eqn:app:rmk:variance}
\text{Var}(T_{\alpha,N}) = \frac{1}{(\lambda^\star)^2}\sum_{i=1}^{N-1}\frac{1}{(i(N-i))^2}.
\end{align}
By using a similar approach as in~(\ref{eqn:app:thm2:increasing mean}), we have
\begin{align*}
& \text{Var}(T_{\alpha,N+1})-\text{Var}(T_{\alpha,N}) \\
&\quad = \frac{1}{(\lambda^\star)^2} \bigg(\frac{1}{N^2}+\sum_{i=1}^{N-1}\frac{1}{(N-i)^2}\Big(\frac{1}{(i+1)^2} - \frac{1}{i^2}\Big)\bigg) \\
&\quad < \frac{1}{(\lambda^\star)^2} \bigg(\frac{1}{N^2}+\frac{1}{(N-1)^2}\sum_{i=1}^{N-1}\Big(\frac{1}{(i+1)^2} - \frac{1}{i^2}\Big)\bigg) \\
&\quad = \frac{1}{(\lambda^\star)^2} \frac{-2}{N^2(N-1)} < 0,
\end{align*}
which proves that the variance of $T_{\alpha,N}$ under the cooperative model is strictly increasing with $N$. To prove the order of $T_{\alpha,N}$, we use the similar approach as in the proof of (T4) in Appendix~\ref{sec:appendix:proof of thm 2}. By noting that (i) $\text{Var}(T_{\alpha,N}) = \frac{2}{(\lambda^\star)^2}\sum_{i=1}^{\frac{N-1}{2}}(f(i))^2$ for an odd $N$ by~(\ref{eqn:app:rmk:variance}), and (ii) $(f(x))^2\,(\deq \frac{1}{(x(N-x))^2})$ is a strictly decreasing convex function, we have
\begin{align*}
\text{Var}(T_{\alpha,N}) \le \frac{2}{(\lambda^\star)^2} \bigg\{ (f(1))^2 + \int_{1}^{\frac{N-1}{2}}(f(x))^2 \,\text{d}x \bigg\}.
\end{align*}
Using basic calculus, we obtain $(f(1))^2 + \int_{1}^{\frac{N-1}{2}}(f(x))^2 \,\text{d}x = \frac{1}{(N-1)^2}+\frac{1}{N^2}-\frac{N+5}{N^2(N^2-1)}+\frac{2}{N^3}\log\frac{(N-1)^2}{N+1}$, which gives $\text{Var}(T_{\alpha,N}) = O((\lambda^\star N)^{-2}).$ By the same reason as above, we have the following lower bound:
\begin{align*}
\text{Var}(T_{\alpha,N}) &\ge \frac{2}{(\lambda^\star)^2} \int_{1}^{\frac{N+1}{2}}(f(x))^2 \,\text{d}x \\
&= \frac{2}{(\lambda^\star)^2}\bigg(\!\frac{1}{N^2}-\frac{N-3}{N^2(N^2-1)}+\frac{2}{N^3}\log(N+1)\!\bigg),
\end{align*}
which gives $\text{Var}(T_{\alpha,N}) = \Omega((\lambda^\star N)^{-2}).$ Hence, we have $\text{Var}(T_{\alpha,N}) = \Theta((\lambda^\star N)^{-2}).$ Similarly as above, we can prove that the skewness of $T_{\alpha,N}$ is also strictly increasing with $N$, and it scales as $\Theta((\lambda^\star N)^{-3})$. Due to similarity, we omit detailed derivations. In our model, the Markov chain $\{\bm{I}_\alpha(t); t \ge 0\}$ is eventually absorbed into the absorbing state space $\mathcal{E}_\alpha^o$ with probability~1, which shows the existence of the finite $n$th moment of $T_{\alpha,N}$~\cite[Eq. (2.2.7)]{neuts81}.

We next consider the non-cooperative model. By independence of $Z_{i}^o$, the variance of $T_{\alpha,N}^o$ is obtained by $\text{Var}(T_{\alpha,N}^o) = \frac{1}{(\lambda^\star)^2}\sum_{i=1}^{N-1}\frac{1}{i^2}.$ Hence, it is strictly decreasing with $N$ and converges to $(\lambda^\star)^{-2}\zeta(2)$ as $N$ goes to~$\infty$. Similarly as above, we can prove that the skewness of $T_{\alpha,N}^o$ is strictly increasing with~$N$ and converges to $(\lambda^\star)^{-3}\zeta(3)$ as $N$ goes to~$\infty$. Due to similarity, we omit detailed derivations. Since $\E[T_{\alpha,N}^o] = \infty $, all the other $n$th moments for $n=2,3,\ldots$ are also divergent as shown below:
\begin{align*}
\E[(T_{\alpha,N}^o)^n] &= \int_{0}^{\infty} \!\! x^n \text{d} \pr\{T_{\alpha,N}^o \le x\} \ge \int_{1}^{\infty} \!\! x^n \text{d} \pr\{T_{\alpha,N}^o \le x\} \\
&\ge \int_{1}^{\infty} x \text{d} \pr\{T_{\alpha,N}^o \le x\} \ge \E[T_{\alpha,N}^o]-1 = \infty.
\end{align*}

\section{Proof of Theorem~\ref{thm:impact of multiple community}}\label{sec:appendix:proof of thm 3}
\begin{figure}[t!]
\centering
{\epsfig{figure=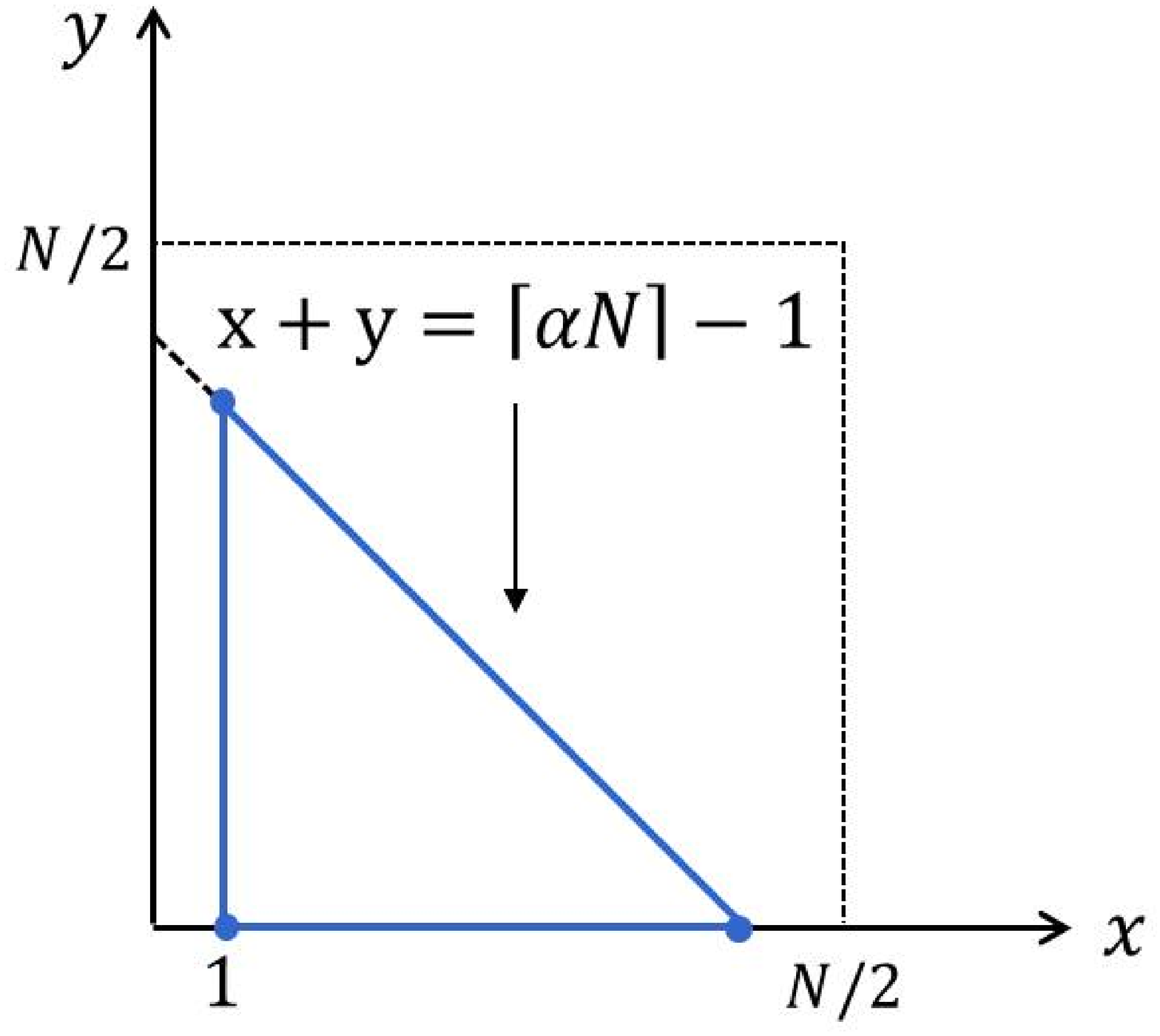,width=0.15\textwidth}}
{\epsfig{figure=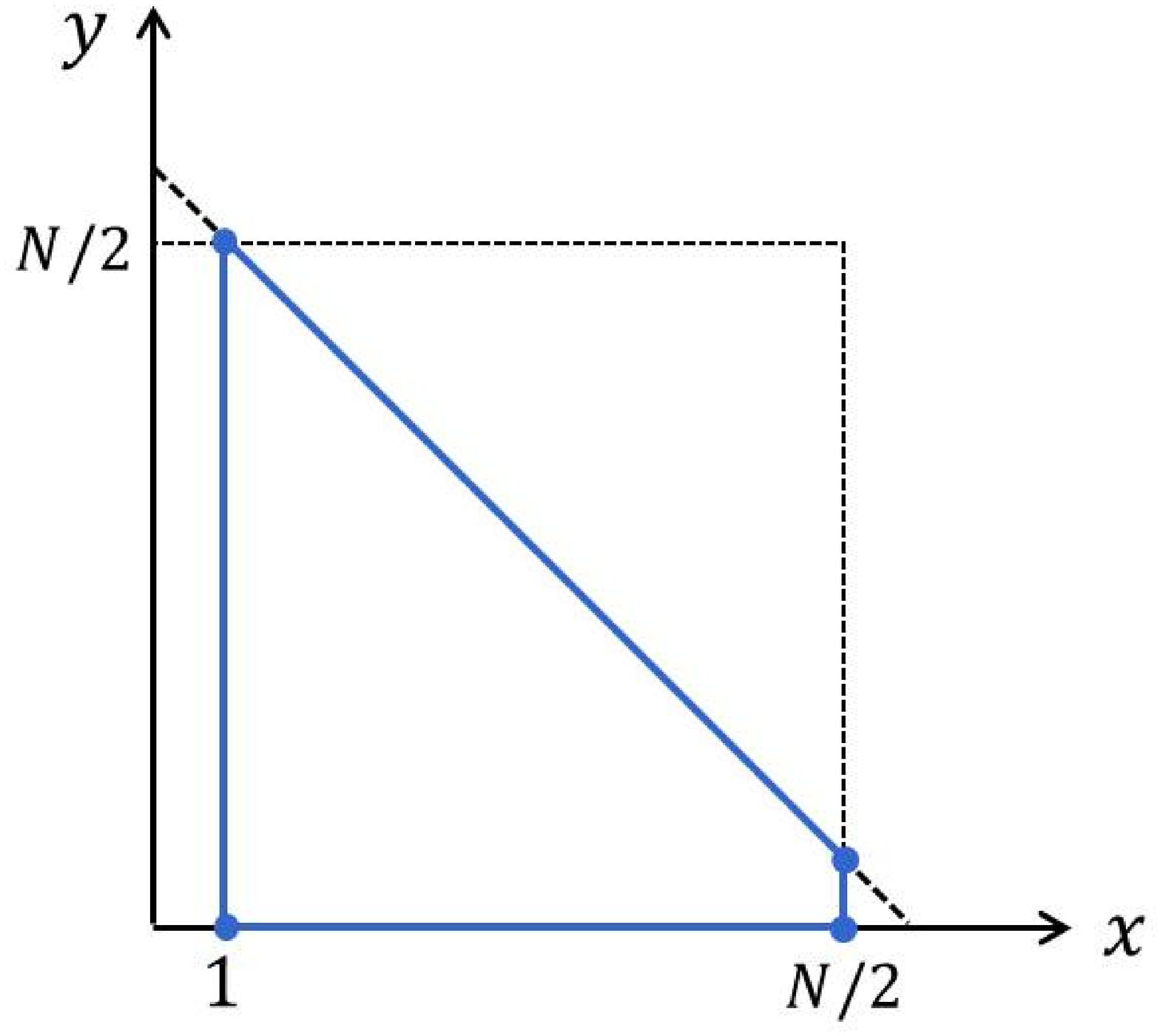,width=0.15\textwidth}}
{\epsfig{figure=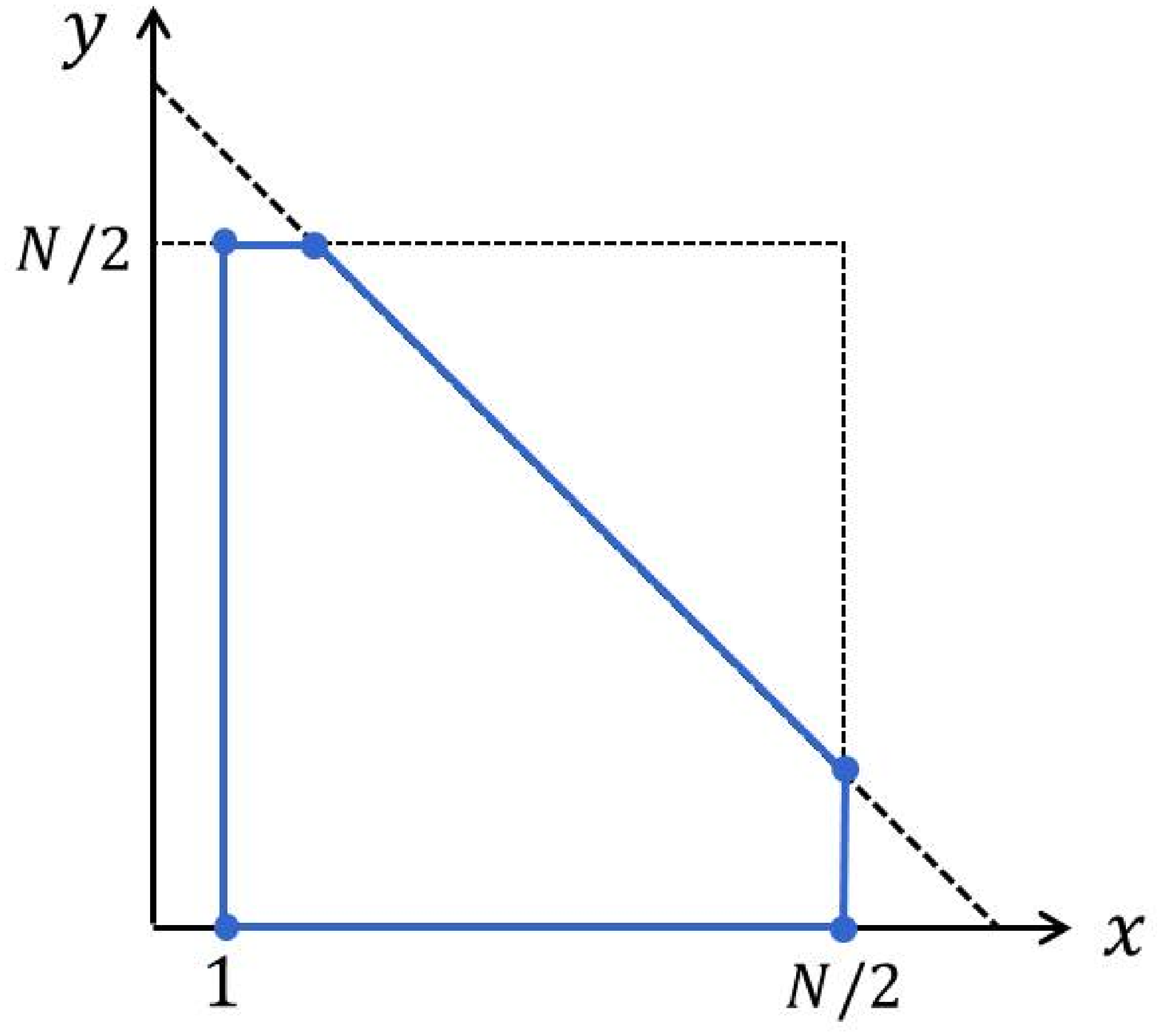,width=0.15\textwidth}}
\caption{Domain $\mathcal{E}_\alpha^\star$ of the maximization problem in~(\ref{eqn:app:thm3:max}) when $\lceil \alpha N\rceil -2 <\frac{N}{2}$ (left), $\lceil \alpha N\rceil -2 =\frac{N}{2}$ (middle), and $\lceil \alpha N\rceil -2 >\frac{N}{2}$ (right).}
\label{fig:domain}
\end{figure}

Without loss of generality, we assume $\lambda_{1,1}^\star \ge \lambda_{2,2}^\star$. Then, by the condition~(\ref{eqn:constraint}) the infection rates can be written in terms of $\lambda^\star$ and $\gamma$ as follows:
\begin{align}\label{eqn:app:thm3:rate}
\begin{split}
\lambda_{1,1}^\star &= \frac{2\gamma\lambda^\star}{\gamma+1}, \\
\lambda_{2,2}^\star &= \frac{2\lambda^\star}{\gamma+1}, \\
\lambda_{1,2}^\star &= \lambda_{2,1}^\star = \lambda^\star.
\end{split}
\end{align}
By the second formula for $H_\alpha(\cdot)$ in Lemma~\ref{lemma:distribution of completion time}, we have $D_{\alpha}(\gamma) = -\max_{i}\rho_i (= \min_i|\rho_i|)$. As shown in Lemma~\ref{lemma:distribution of completion time}, diagonal elements of $\bm{F}_\alpha$ constitute $\rho_i$, which are negative of transition rate from $(i_1,i_2)$ to the set $\{(i_1+1,i_2),(i_1,i_2+1)\}$ for $(i_1,i_2)\in\mathcal{E}_\alpha^\star$. Hence, we can obtain $D_{\alpha}(\gamma)$ by solving the following maximization problem:
\begin{align}\label{eqn:app:thm3:max}
D_{\alpha}(\gamma) = -\max_{(i_1,i_2)\in\mathcal{E}_\alpha^\star} \rho(i_1,i_2),
\end{align}
where
\begin{subequations}
\begin{align}
\rho(i_1,i_2) &= -\sum_{l=1}^{2}\Big(\frac{N}{2}-i_l\Big)\sum_{k=1}^{2}(i_k \lambda_{k,l}^\star), \label{eqn:app:thm3:max(1)} \\
\mathcal{E}_\alpha^\star &= \{(i_1,i_2)\in\mathbb{Z}^2_+\,|\,
1 \le i_1+i_2 \le \lceil \alpha N \rceil-1, \nonumber \label{eqn:app:thm3:max(2)} \\
& \hspace{7 mm} 1\le i_1 \le N/2, 0 \le i_2 \le N/2\}.
\end{align}
\end{subequations}
Note that
\begin{align*}
& \frac{\partial^2 \rho(i_1,i_2)}{\partial (i_1)^2} = 2\lambda_{1,1}^\star >0, \\
& \frac{\partial^2 \rho(i_1,i_2)}{\partial (i_2)^2} = 2\lambda_{2,2}^\star >0, \\
& \frac{\partial^2 \rho(i_1,\lceil \alpha N \rceil -1-i_1)}{\partial (i_1)^2} = 0,
\end{align*}
which implies that the maximum of $\rho(i_1,i_2)$ in~(\ref{eqn:app:thm3:max}) occurs at the vertex of the set $\mathcal{E}_\alpha^\star$ in~(\ref{eqn:app:thm3:max(2)}) (see Fig.~\ref{fig:domain}). Suppose $\lceil \alpha N\rceil -2 <N/2$. Then, from the figure, the vertices of $\mathcal{E}_\alpha^\star$ are given by $(1,0), (1,\lceil \alpha N \rceil-2)$, and $(\lceil \alpha N \rceil-1,0)$, and thus we have
\begin{align*}
&\max_{(i_1,i_2)\in\mathcal{E}_\alpha^\star} \rho(i_1,i_2) \\
&\quad = \max\{\rho(1,0),\rho(1,\lceil \alpha N \rceil-2),\rho(\lceil \alpha N \rceil-1,0)\}  \\
&\quad = \rho(1,0).
\end{align*}
Similarly as above, we can solve the maximization problem for each of the cases when $\lceil \alpha N\rceil -2 =N/2$ and $\lceil \alpha N\rceil -2 >N/2$. Since it is straightforward, we summarize the results without detailed derivations:
\begin{align*}
&\max_{(i_1,i_2)\in\mathcal{E}_\alpha^\star} \rho(i_1,i_2) \\
&=
\begin{cases}
\rho(1,0), & \alpha\le 1-\frac{2}{N}, \\
\rho(\frac{N}{2},\lceil \alpha N\rceil -1 -\frac{N}{2}), & \alpha=1, \\
\rho(1,0), & 1-\frac{2}{N} <\alpha <1, \gamma < \frac{5N-16}{N-4}, \\
\rho(\frac{N}{2},\lceil \alpha N\rceil -1 -\frac{N}{2}), & 1-\frac{2}{N} <\alpha <1, \gamma > \frac{5N-16}{N-4}. \\
\end{cases}
\end{align*}
From (\ref{eqn:app:thm3:rate}) and (\ref{eqn:app:thm3:max(1)}), we have
\begin{align*}
\rho(1,0) &= -\frac{(N-2)\gamma\lambda^\star}{\gamma+1}-\frac{N\lambda^\star}{2}, \\
\rho\Big(\frac{N}{2},\lceil \alpha N\rceil -1 -\frac{N}{2}\Big) &= -(N-\lceil \alpha N\rceil +1)\bigg\{\frac{N\lambda^\star}{2} \\
& \qquad +\Big(\lceil \alpha N\rceil-1-\frac{N}{2}\Big)\frac{2\lambda^\star}{\gamma+1}\bigg\}.
\end{align*}
Therefore, $\frac{\text{d}}{\text{d}\gamma}\rho(1,0)<0$ and $\frac{\text{d}}{\text{d}\gamma}\rho(\frac{N}{2},\lceil \alpha N\rceil -1 -\frac{N}{2}) >0$ for all $\gamma \ge 1$, which proves the theorem.

\end{document}